\documentclass[aps,pre,twocolumn,showpacs]{revtex4-1}
\usepackage{graphicx,color}
\usepackage{amsmath,amssymb}

\begin{document}
\title{Regular modes of a mixed dynamical based optical fiber }
\author{C. Michel,$^1$ M. Allgaier$^2$ and V. Doya$^1$\\}
\affiliation{$^1$ Laboratoire de Physique de la Mati\`ere Condens\'ee, CNRS UMR 7336, Universit\'e Nice-Sophia Antipolis, 06100 Nice, France\\
$^2$ Fachbereich Physik der Philipps-Universit\"at Marburg, D-35032 Marburg, Germany}

\begin{abstract}
A multimode optical fiber with a truncated transverse cross section acts as a powerful versatile support to investigate the wave features of complex ray dynamics. In this paper, we concentrate on the case of a geometry inducing a mixed dynamics. We highlight the properties of regular modes associated to stable periodic orbits such as an enhanced intensity localization and report unexpected attributes such as the statistics of the Inverse Participation Ratio that present features analogous to those of Anderson localized modes. Our study is supported by both numerical and experimental results. 
\end{abstract}


\maketitle


\section{Introduction}
Chaotic systems are more and more used in applications. These systems, whose geometry induces  complexity through a chaotic ray dynamics, bring a new kind of solutions for practical purposes.
Recent developments in the field of electromagnetic compatibility show that chaotic geometries promote a very efficient mode mixing in electromagnetic chambers \cite{Gros2014,Selemani2015}. In optics, dielectric microcavities stand as the typical example of the exploitation of chaos as an improvement of existing devices \cite{Cao2015}. Indeed, chaotic microcavity lasers have shown to allow low threshold and highly directional outputs \cite{Gmachl1998,Harayama2003b} and are a very useful solution for integrated optics. Recently, microcavity-based gyroscopes in which a chaotic cavity enhances the rotation sensitivity \cite{Ge2015} have been investigated. Before that, a chaotic double-clad fiber amplifier had been proven to be an optimized solution for a uniform pump absorption along the total amplifier length \cite{Doya2001}. 

Instead of considering the fully chaotic regime, one can consider a so-called \textsl{mixed} system. In this case, the coexistence of regular and chaotic motions leads to an enlarged wealth of the dynamics. This duality gives for instance rise to a wave feature known as dynamical tunneling, which allows a passage between the chaotic and regular regions by coupling one to another \cite{Backer2009}. This effect has been used in segmented waveguides where tunneling between stability islands ensures a non-diffractive beam regime \cite{Aschieri2013}, as well as in deformed microdisks in which light is coupled to a bus waveguide through the tunneling effect acting as a resonant dynamical filter \cite{Song2014}.

A highly multimode optical fiber whose transverse cross-section presents a truncated-circle shape allows to explore different regimes of the dynamics by changing the length of the truncated diameter (\textsl{e.g.} changing $d$ in Fig. \ref{Presentation}(b)). Over the past few years we studied experimentally manifestations of ergodicity as well as deviations from this universal behavior  due to scar modes in passive and active chaotic optical fibers \cite{Doya2002,Doya2001a,Michel2009a,Michel2012}. We keep on exploiting the potential of multimode fibers, but now concentrating studies on wave features of mixed systems. The latter present both regular and chaotic dynamics in the ray limit, as exemplified by the dynamics of the so-called Mushroom billiard \cite{Backer2008}. In this paper, we focus on the properties of the modes of a mixed dynamics based multimode optical fiber, and in particular modes that are associated to the regular part of the dynamics. Unlike ergodic modes whose statistics follow a universal behavior, the \textsl{regular modes} exhibit some strong deviations that are unexpectedly analogous to those induced by localized modes in disordered systems \cite{Lagendijk2009}.

In section \ref{SecPresentation}, we present our experimental system. We also present two usual manners of distinguishing regular, chaotic and mixed systems, on the one hand through their Poincar\'e Surface of Section and on the other hand through their spectral statistics. In section \ref{Resonator}, we develop the analysis of the spatial properties of the regular modes of the fiber using the analogy with the gaussian modes of an optical resonator. In section \ref{Localisation} we propose an original point of view by interpreting our results through tools usually used in disordered systems exhibiting (Anderson) localized modes. In section \ref{Experimental}, we present an analysis of the properties of the regular modes of the optical fiber based on both experimental and numerical results. Then, in section \ref{Conclusion}, we conclude and propose, in an open discussion, some potential applications.


\section{The optical fiber as a versatile optical analogue of a dynamical system}
\label{SecPresentation}

Our experimental system is a non-standard silica multimode optical fiber, whose cross section is a truncated circle. The diameter of the core of the fiber is $2a=125\mu$m, and the truncated diameter $d=\gamma\times a$, with $\gamma \in \mathopen{]}0\,;2\mathclose{]}$. 
\begin{figure}
\centering
\includegraphics[width=\columnwidth]{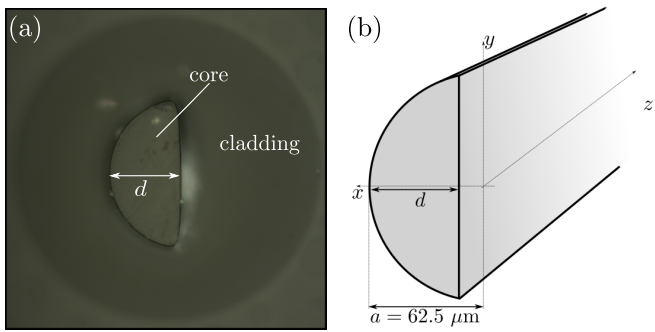}
\caption{Presentation of the experimental system. (a) picture of the cross section of the optical fiber for $\gamma=0.95$ taken with a standard microscope in the transmission mode, (b) scheme of the fiber with $\gamma<1$.}
\label{Presentation}
\end{figure}
The core is surrounded by a silicon cladding of diameter $250\mu$m  [Fig. \ref{Presentation}]. The optical indexes are $n_{\rm co}=1.458$ and $n_{\rm cl}=1.41$ for the core and the cladding respectively for the vacuum wavelength $\lambda_0=632$nm. As the wavelength is small compared with the characteristic size $a$ of our system, the semi-classical approximation is valid, and one can consider both the classical description of ray dynamics as well as a modal description of propagative waves. 
\begin{figure*}
\centering
\includegraphics[width=\textwidth]{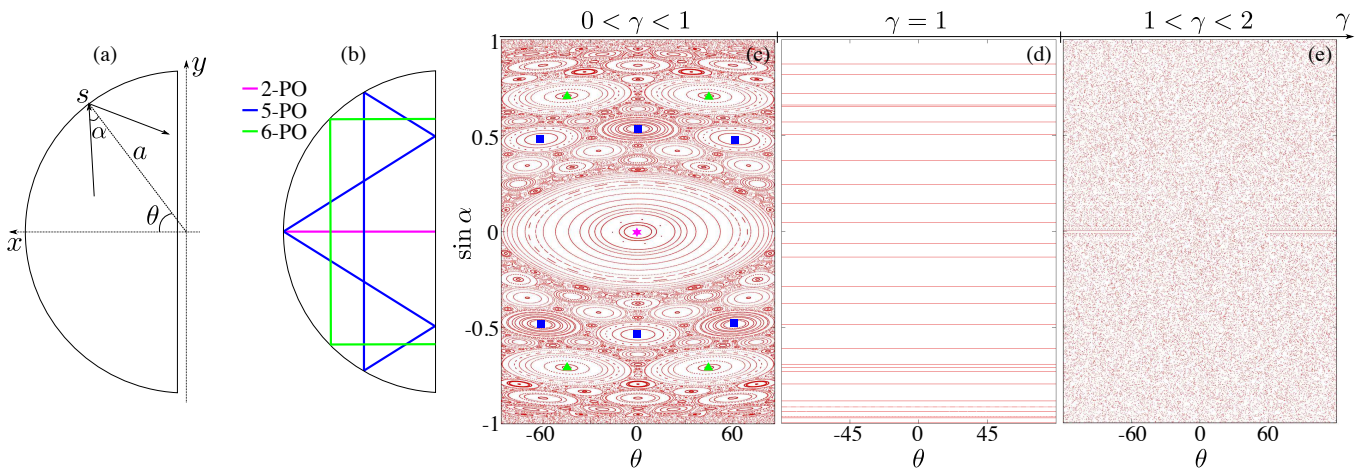}
\caption{(a) Curvilinear abscissa $s=a\theta$ and angle of reflection $\alpha$ used for the calculation of the  SOS and Husimi representations. (b) Three examples of stable periodic orbits of the mixed billiard ($\gamma=0.95$) : in pink the 2-PO, in blue the 5-PO and in green the 6-PO.(c)-(e) Poincar\'e Surface of Section representation for three different values of $\gamma$. (c) $\gamma=0.95$, mixed dynamics with a coexistence of regular and chaotic trajectories ; the pink star points out the 2-bounce stable periodic orbit, the blue squares correspond to the central point of the stability islands of the 6-bounce stable periodic orbit and the green triangles to the 5-bounce stable periodic orbit, (d) $\gamma=1$, regular dynamics for which one sees the conservation of the angle of reflection, whatever the initial condition is, (e) $\gamma=2/3$, chaotic dynamics, no stability regions, all the periodic orbits are unstable.}
\label{TracePSS}
\end{figure*}
In this regime, the cross-section of the optical fiber is equivalent to a two-dimensional (2D) bounded cavity which is a well-known paradigm of a hamiltonian dynamical system (2D billiard) to study wave chaos \cite{Stockmann1999}. Indeed, the  longitudinal evolution of a ray along the fiber is equivalent to the time evolution of a trajectory in a billiard. Thus, in the following, we will refer to the transverse cross section of the optical fiber as a ``cavity''.

The versatility of our system relies on its ability to explore different types of dynamics from regular to chaotic by varying the factor $\gamma$. The qualification of the dynamics is commonly studied through a stroboscopic phase-space representation, well-known as the Poincar\'e Surface of Section (SOS). The SOS is a reduction of the ray dynamics to a 2-dimensional mapping.  It consists in plotting at each impact of the ray on the boundary the curvilinear abscissa $s$ measured through the angle $\theta=s/a$ and the sine of the incidence angle, $\alpha$ [Fig. \ref{TracePSS}(a)].
As shown in Fig. \ref{TracePSS}(c-e) and as pointed out by Ree \cite{Ree1999}, the SOS presents the characteristic behaviors related to different types of dynamics while varying $\gamma$. The nature of the classical dynamics in a D-shaped billiard as been proved to depend on the stability of the shortest periodic orbit (2-bounce periodic orbit (2-PO)) along the diameter $d$. When $0<\gamma<1$, the truncated diameter $d$ is lower than the radius of the billiard. In such a configuration, the dynamics is mixed and the SOS presents a complex combination of stochastic and regular regions. In figure [Fig. \ref{TracePSS}(c)] plotted for $\gamma=0.95$, regular islands coexist with diffuse points associated to the so-called chaotic sea \cite{Tureci2002}. Each regular island corresponds to quasi-periodic trajectory in the vicinity of a stable periodic trajectory at the center of the island. These stability islands mark the regular dynamics while the chaotic sea is a signature of chaotic trajectories. A large fraction of SOS is occupied by the main central resonance that corresponds to the 2-PO which is stable in this case.
As $d$ reaches the value of the radius $a$ ($\gamma=1$, half-circle), the SOS testifies of the regular nature of the dynamics [Fig. \ref{TracePSS}(d)]. In this case, the angle of reflection $\alpha$ remains indefinitely the same for a given initial condition. Note that the SOS of a circular billiard ($\gamma=2$) would be the same due to angle conservation. When $d$ becomes greater than $a$ ($1<\gamma<2$), no more regular islands exist, and all the periodic orbits (PO) become unstable. Then, the SOS is densely covered by diffuse points and the dynamics is chaotic [Fig. \ref{TracePSS}(e)].
It is worth mentioning that the latter geometry has also been fully studied in optical fibers over the past few years \cite{Doya2002,Doya2001a}. In particular, we experimentally demonstrated that some specific scarred modes of a chaotic optical fiber -- spatially localized along the least unstable periodic orbits -- can be selectively enhanced through an optical amplification process \cite{Michel2007,Michel2012}. Moreover, as the pump absorption is clearly improved by the use of a chaotic double-clad fiber amplifier, the device has been proposed as an optimization of existing amplifiers \cite{Doya2001,Roska}.

\begin{figure}
\centering
\includegraphics[width=\columnwidth]{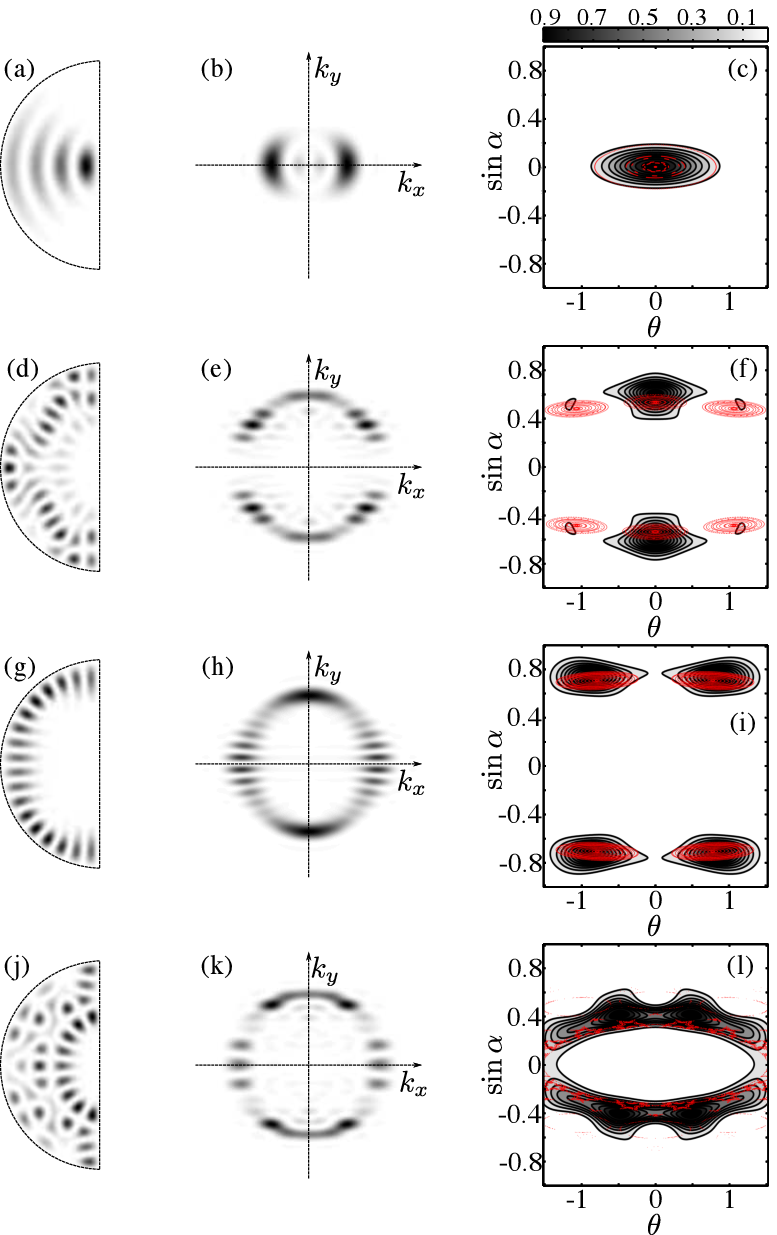}
\caption{Near-field, Far-field and Husimi representation of (a,b,c) a regular mode of the 2-PO. Super-imposed in red is the SOS associated to the 2-PO ; (d,e,f) a mode built on the 5-PO. In red, the SOS of the 5-PO ; (g,h,i) a mode built on the 5-PO. In red the SOS of the 5-PO ; (j,k,l) a chaotic mode.}
\label{Modes}
\end{figure}

We now focus our study on the system with a mixed dynamics, and thus consider the case $\gamma=0.95$ which corresponds to our fabricated fiber. The length of the fiber is 10 cm, which is greater than the Heisenberg length \cite{Doya2002} ensuring the validity of a modal description of light propagation. The number of modes of the fiber is given by the standard formula \cite{Ghatak1998} and is evaluated to approximatively $N=6000$ at $\lambda_0$. As shown on four exemples in the Figure \ref{Modes}, the modes exhibit some signatures of the underlying classical mixed dynamics \cite{Percival1973}. The spatial distribution of intensity (near-field, NF) of the  mode in Fig. \ref{Modes}(a) is located on the 2-PO represented in pink in Fig. \ref{TracePSS}(b). As mentioned before, this 2-PO corresponds to the central point of the major stability island in the SOS [Fig. \ref{TracePSS}(c)]. Figure \ref{Modes}(b) shows the far-field (FF) of the same mode. The FF is the square modulus of the spatial Fourier transform of the field distribution. It indicates the direction $\vec{\kappa}$ and modulus $\kappa$ of the transverse wave vectors. For Fig. \ref{Modes}(b), two maxima of intensity are localized on the $\kappa_x$-direction of the 2-PO at a distance $\kappa$ of the center of the Fourier space. This directivity in the FF evidences the signature of the underlying 2-PO (see Fig. \ref{TracePSS}).
Strong correspondances between some particular modes and their associated classical trajectories in the semiclassical regime are established through the Husimi representation, which is commonly used as a wave equivalent to the classical SOS \cite{Husimi1940}. In the case of 2D cavities, the Husimi function measures the normal derivative of the eigenfunction on the boundaries of the cavity \cite{Birkhoff1927,Crespi1993,Ree1999,Backer2004}. Here, the Husimi function is calculated for the modes of the optical fiber and is represented in the 2D space $(\theta,\sin{\alpha})$. Here, $\theta$ is the curvilinear abscissa normalized to the radius as described in Figure \ref{TracePSS}(a) ; $\sin\alpha$ is the projection of the transverse wave vector $\vec \kappa$ along the tangential direction to the boundary, normalized by the modulus $\kappa=|\vec \kappa|$ of the mode. Figure \ref{Modes}(c) presents the Husimi representation of the mode shown in Fig. \ref{Modes}(a) with the SOS of the 2-PO superimposed. The localized pattern of the intensity in the Husimi representation means that the mode is strongly confined along the $x$-direction. This confinement as well as the very good agreement between both this representation and the SOS confirms that this mode builds on constructive interferences along the 2-PO. In the same way, Figures \ref{Modes}(d,e,f) and (g,h,i) show the NF, FF and Husimi representations of modes built in the vicinity of the 5-bounce [Fig. \ref{TracePSS}(b), blue line] and 6-bounce [Fig. \ref{TracePSS}(b), green line] stable periodic orbits respectively. As before, the agreement between the Husimi and SOS representation reflects the fact that the ray dynamics underlies the modal behavior. Figures \ref{Modes}(j,k,l) present the NF, FF and Husimi representations of a chaotic mode of the fiber. The main difference is highlighted by the Husimi representation which is no longer confined in small areas delimited by the corresponding stability islands of the SOS, but spread over the chaotic regions.

\begin{figure}
\centering
\includegraphics[width=\columnwidth]{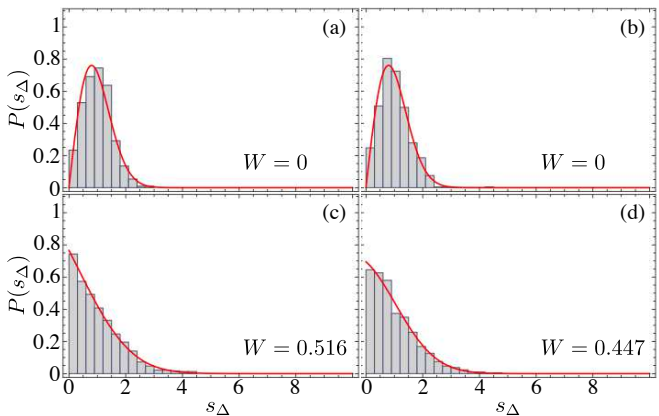}
\caption{Probability distribution of the energy level spacing for the symmetric and anti-symmetric modes of (a,b) a chaotic cavity ($\gamma=3/2$) and (c,d) a mixed cavity ($\gamma=0.95$).}
\label{Hist}
\end{figure}

As conjectured by Percival \cite{Percival1973}, the duality between chaotic and regular dynamics is also encountered in the spectral features. A simple way to quantify the degree of regularity of a 2D-cavity is given by the study of the statistics of energy level spacing 
\begin{equation}
	s_\Delta=\frac{E_{n+1}-E_n}{\frac{1}{N}\sum_{n=1}^N{E_n}}
\end{equation}
where $E_n$ is the energy of mode $n$ and $N$ the total number of modes. As described in \cite{Berry1984,Rudolf2012}, the probability distribution $P\left(s_\Delta\right)$ in a mixed system is given by
\begin{equation}
	P(s_\Delta)=\frac{d^2}{ds_\Delta^2	}\left[ \exp{\left( -Ws_\Delta \right)} \text{ erfc}\left( \frac{\sqrt{\pi}}{2}\left( 1-W \right)s_\Delta\right) \right]
\end{equation}
where $0<W<1$ is the density of regular states. A formal evaluation of $W$  is deduced directly from the SOS, by measuring the area of the regular region, weighted by the lengths of the associated trajectories. Figure \ref{Hist} shows the statistics of a chaotic cavity ($\gamma=3/2$) for symmetric [Fig. \ref{Hist}(a)] and antisymmetric [Fig. \ref{Hist}(b)] modes and a mixed cavity ($\gamma=0.95$) for symmetric [Fig. \ref{Hist}(c)] and antisymmetric [Fig. \ref{Hist}(d)] modes. The gray scale histograms represent the energy level spacing distribution calculated for the modes of both cavities and the red line is a fit of the histograms by the function $P(s_\Delta)$ giving an estimation of the parameter $W$. For the chaotic case, $W=0$ indicates that the relative size of the regular region is reduced to zero. This result is consistent with the typical SOS of a chaotic system [Fig. \ref{TracePSS}(e)]. Indeed, the inherent nature of a chaotic dynamics is to avoid any stable structure.
On the contrary, for the mixed system, $W\simeq 0.5$ corresponds to a SOS half-filled with the regular region. Note that this is qualitatively coherent with the SOS presented in Fig. \ref{TracePSS}(c). The case $W=1$ would correspond to regular systems $\gamma=1$ (half-circle) and $\gamma=2$ (circle).


\section{Regular modes as modes of a plano-concave resonator}
\label{Resonator}

We now go one step further in the analogy between longitudinal optical ray propagation in the fiber and time evolution of the ray in the billiard (here the transverse cross section of the fiber).
As evidenced in details by H. E. Turecci in \cite{Tureci2002} in dielectric microcavities, the modes localized on stable POs can be described in terms of gaussian optical theory.
Using the standard physics of optical resonators, we develop that correspondence for the 2-PO regular modes as the analogous of the fundamental longitudinal modes of a plano-concave optical resonator with dielectric boundaries where $x$ acts as the longitudinal axis of propagation and $y$ as the transverse direction. Thus, the optical resonator formalism allows us to analyze both the stability of the cavity and the spatial field distribution of the modes of the fiber.

First, the ray transfer matrix formalism enables to study the stability of the cavity by investigating the paraxial ray evolution in the vicinity of the 2-PO. Ray transfer matrix theory gives the state of output quantities $s_{out}$ and $\alpha_{out}$ in terms of input quantities $s_{in}$ and $\alpha_{in}$ using the optical characteristics of the resonator and its boundary conditions. Here, $s$ refers, as described earlier, to the position of the beam on the resonator boundary and $\alpha$ to the angle of the beam compared with the normal on the boundary. In the paraxial approximation, output quantities are linearly dependent on input quantities :
\begin{equation}
\begin{pmatrix}
s_{out}\\
\alpha_{out} 
\end{pmatrix}
=M_t
\begin{pmatrix}
s_{in}\\
\alpha_{in}
\end{pmatrix}
\end{equation} 
where $M_t$ is the ray transfer matrix. The stability of the resonator is given by the value of the trace of $M_t$ associated to a periodic sequence of a given ray. When rays are periodically refocused, the sequence is stable and geometric characteristics of the resonator obey the relation:
\begin{equation}
0\leq\left(1-\frac{d}{R_1}\right)\left(1-\frac{d}{R_2}\right)\leq1\label{stability}
\end{equation}
where $d=\gamma\times a$ is the resonator length and $R_1$, $R_2$ are the radii of curvature of the resonator boundaries. This stability condition is analogous to the stability analysis of dynamical systems based on the evolution of a small deviation in the vicinity of a PO \cite{Berry1983}. The transfer matrix is then formally equivalent to the so-called monodromy matrix.

In our case, $R_1=a$ and $R_2=\infty$ because of the flat reflective boundary. From the relation (\ref{stability}), one sees that the stability of the resonator depends on the value of $d$.  Cases $\gamma=1$ and $2$ are respectively associated to the hemispheric and confocal stable resonators. For $1<\gamma<2$, the resonator is unstable. For $\gamma<1$, the condition (\ref{stability}) is fulfilled so that the resonator is stable.
The mixed fiber with $\gamma=0.95$ can thus be considered as a stable optical resonator whereas the chaotic fiber ($\gamma=3/2$) corresponds to an unstable resonator \cite{Ree1999,Bogomolny1988,Berry1984}. As a matter of fact, the nature of the dynamics of the underlying billiard with a truncated circle shape is conditioned by the stability of the 2-PO as mentioned before.

\begin{figure}
\centering
\includegraphics[width=\columnwidth]{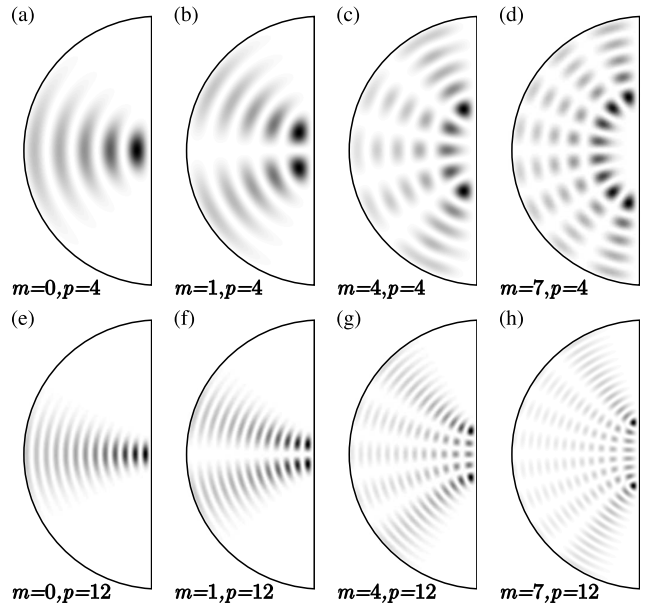}
\caption{Modes of the mixed fiber associated to different values of $(m,p)$. (a)-(d) $p=4$ and $m=0,1,4,7$, (e)-(h) $p=12$ and $m=0,1,4,7$.}
\label{modes}
\end{figure}

In the following, we consider the case of the stable resonator with $\gamma=0.95$, corresponding to the cross section of our optical fiber. 
We use the gaussian wave formalism in the $(x,y)$-plane in order to study the spatial distribution of the modes built on the 2-PO. The spatial expansion of a beam traveling in the $x$-direction in the slowly varying envelope approximation reads:
\begin{equation}
\psi(x,y)=u(x,y)\times\exp{(-j\kappa x)}
\label{psi}
\end{equation}   
where $u(x,y)$ is a slowly varying function describing the deformation of a plane wave due to the curvatures of the resonator walls and $\psi(x,y)$ is a solution of the transverse Helmholtz scalar equation $\Delta\psi+\kappa^2\psi=0$. By using expression (\ref{psi}) for $\psi(x,y)$, one gets the following Fresnel (or parabolic) form for the wave equation in the paraxial approximation:
\begin{equation}
\frac{\partial^2 u}{\partial x^2} + \frac{\partial^2 u}{\partial y^2}-2j\kappa\frac{\partial u}{\partial x}=0\label{fresnel}
\end{equation} 
The solution $u(x,y)$ of (\ref{fresnel}) -- under the assumption of $u(x,y)$ varying slowly enough with $x$ so that $\partial^2u/\partial^2x\ll |2\kappa\partial u/\partial x|$ and under the condition of a matching between the wave front and the curved and plane walls of the resonator -- is given by \cite{Kogelnik1966}:
\begin{eqnarray}
u(x,y)&=&\frac{w_0}{w(x)}H_m\left(\sqrt{2}\frac{y}{w(x)}\right)\label{u}\\
&\times &\exp\left\{-j(\kappa x-\phi_{m,p})-y^2\left(\frac{1}{w^2(x)}+j\frac{\kappa}{2R(x)}\right)\right\}\nonumber
\end{eqnarray}
where $w(x)$ is the beam radius, $w_0$ is the minimum beam diameter -- the so-called waist diameter -- for which the phase front is plane, $R(x)$ the radius of curvature of the wavefront and $H_m$ is the Hermite function of order $m$. The solution $u(x,y)$ is consistant with the theory of gaussian beams along the $y$-direction \cite{Kogelnik1966}. The gaussian beam profile is characterized by the beam radius $w(x)$.
An initial gaussian beam of width $w_0$ at $x=0$ experiences a transverse expansion given by $w^2(x)=w_0^2\left(1+\left(x/x_{\rm r}\right)^2\right)$ during its propagation along the $x$-axis. $x_{\rm r}$ is the Rayleigh length, measuring the spatial coherence of the beam along the axis of propagation $x$. In our specific case, the relevant wavelength is $\lambda_{\perp}=2\pi/\kappa$. Then, in our case, the Rayleigh length $x_{\rm r}$ and the beam waist $w_0$ are linked by the relation $x_{\rm r}=\kappa w_0^2/2$. The beam propagating along the 2-PO also undergoes a phase shift  which is twofold. First, the reflection on the dielectric core/cladding interface -- governed by Fresnel reflections laws -- implies a phase shift $\phi_{r}^{\kappa}$ which depends on $\kappa$. Second, the Gouy phase $\phi_{g}$ \cite{Feng2001,Kogelnik1966} appears when the beam focuses. It results a complex total phase shift $\phi_t$ that reads :
\begin{eqnarray}
\phi_{t}&=&\phi_{r}^{\kappa}+\phi_{g}\label{phase}\\
&=&2\arctan\sqrt{\left(n^2_{\rm co}-n^2_{\rm cl}\right)k^2_0/\kappa^2-1}\nonumber\\
&+&(m+1)\arctan\sqrt{\frac{\gamma}{1-\gamma}}\nonumber
\end{eqnarray}
where $k_0=2\pi/\lambda_0$ is the modulus of the vacuum wavevector. Stationary modes occur when the accumulated phase shift along a round-trip in the resonator is a multiple of $2\pi$. From (\ref{u}) and (\ref{phase}), this condition of constructive interferences leads to:
\begin{equation}
\kappa_{m,p}\times 2d=2\pi p+2\phi_{r}^\kappa+2(m+1)\arctan\sqrt{\frac{\gamma}{1-\gamma}}\label{kt}
\end{equation}
Each mode of the resonator is then defined by the value of its transverse wave vector $\kappa_{m,p}$ which depends on two integers, $m$ and $p$ running from 0 to maximum values satisfying $\kappa_{m,p}<k_0\sqrt{\left(n^2_{\rm co}-n^2_{\rm cl}\right)}$ \cite{Doya2002}. In the optical resonator analogy, $p$ defines the number of nodes along the axial direction $x$, that is the order of the longitudinal mode along the 2-PO whereas $m$ is the transverse mode number associated to the so-called high-order modes. The fundamental gaussian mode, that is the fundamental mode for the transverse oscillations, corresponds to $m=0$ and $p=0$. Some examples of the modes of the fiber/resonator with their corresponding value of $(m,p)$ are given in Figure \ref{modes}.


\section{Regular modes as localized modes induced by the mixed dynamics}
\label{Localisation}
\begin{figure}
\centering
\includegraphics[width=\columnwidth]{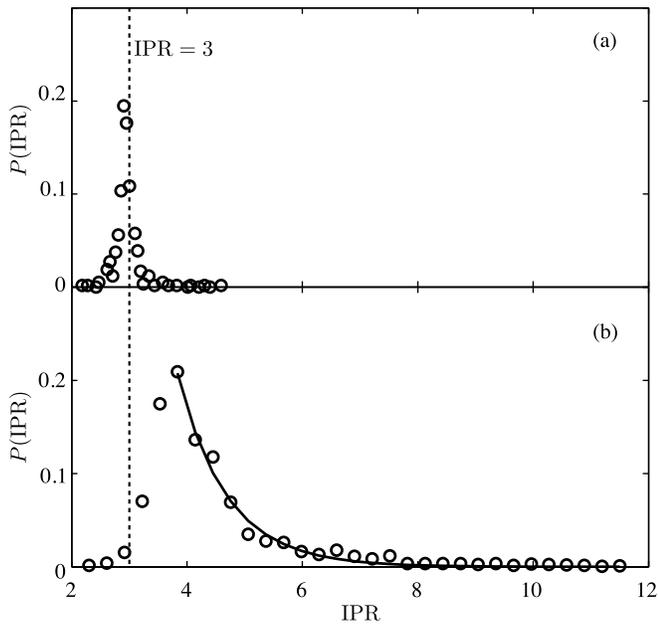}
\caption{Probability distribution of the Inverse Participation Ratio. (a) Chaotic cavity ($\gamma=3/2$). The probability distribution is centered around the value IPR=3. (b) Mixed cavity ($\gamma=0.95$) presenting some larger values of the IPR. The continuous line is a fit based on a nonlinear  sigma model \cite{Pradhan2000}.}
\label{distributionIPR}
\end{figure}

The ergodic modes of a chaotic system stand for the generic behavior. They are characterized by a gaussian statistic of the spatial field distribution. Nevertheless, some modes, namely seen as an ``\textsl{extra density (that) surrounds the region of the periodic orbit}'' \cite{Heller1984} are constructed along \textsl{unstable} periodic orbits. They are usually named ``scar modes''. Regular modes of a mixed system also result from constructive interferences although they take place along a \textsl{stable} periodic orbit. This intrinsic difference in the nature of the underlying periodic orbit implies, among other features, a strong spatial localization of the field [Fig. \ref{modes}(a) and (e)]. This localization implies a substantial deviation in the statistics compared to the homogeneous field distribution.
The Inverse Participation Ratio (IPR) and its statistics, which are commonly-used tools in the characterization of the spatial signatures of chaotic or disordered systems, notably allow to highlight this deviation. The IPR, that is the second order moment $I_2$ of the intensity, can also be defined as :
\begin{equation}
	I_2=\frac{\iint_\mathcal{S} I^2(x,y)d\mathcal{S}}{\left(\iint_\mathcal{S} I(x,y)d\mathcal{S}\right)^2}
	\label{eqIPR}
\end{equation}
where $I(x,y)=|u(x,y)|^2$ is the field intensity, $\mathcal{S}$ the surface of the transverse cross section of the fiber and $\frac{1}{\mathcal{S}}\iint_{\mathcal{S}}I(x,y)d\mathcal{S}=1$.
The distribution of the IPR [Fig. \ref{distributionIPR}] for the modes of the chaotic and mixed fibers exhibit some specific features. 
$P(I_2)$ for the chaotic fiber ($\gamma=3/2$) presents a peaked distribution around the value $I_2=3$ \cite{Kudrolli1995,Pradhan2002a} as expected by the universal properties of ergodicity of the generic modes of a chaotic system \cite{Doya2002}. Even if scar modes present some enhancement of intensity, this is marginal enough not to appear in the distribution of the IPR [Fig. \ref{distributionIPR}(a)]. The Random Matrix Theory (RMT) predicts that there should be no fluctuations, but as the system is bounded, fluctuations appear \cite{Pradhan2000}.
On the contrary, the distribution of IPR calculated for the modes of the mixed fiber shows an asymmetric profile with enlarged high values up to 2 times the actual mean value $\langle I_2 \rangle \simeq 4.5$. This points out that a large number of modes contributes to this deviation by presenting highly localized intensities.
In disordered systems, this behavior is currently associated to the presence of so-called ``localized modes'' \cite{Lagendijk2009}: some theoretical studies based on the supersymmetry method \cite{Prigodin1998} predict the asymmetry of the IPR distribution in the (Anderson) localization regime. The distribution of the IPR is then expected to follow :
\begin{equation}
	P(I_2)=C\times \sqrt{\frac{g}{I_2}}\times \exp{\left( -\frac{\pi}{6}gI_2 \right)}
	\label{PI2}
\end{equation}
for $I_2>\langle I_2\rangle$ where $\langle I_2\rangle$ is the mean-value of the IPR, $C$ is a normalization constant and $g$ the conductivity of the system that depends on the system size, $\langle I_2\rangle$, and the mean-free-path $l$ \cite{Pradhan2000}. The high values of $I_2$ in Fig. \ref{distributionIPR}(b) present a very good agreement with eq. (\ref{PI2}) with $g=2.02$. This value is associated to a mean-free-path $l=16.8\mu$m. The mean-free-path here is smaller than the typical size of the system whereas it is expected to approach the size of the system in the ergodic case.

\begin{figure}
\centering
\includegraphics[width=\columnwidth]{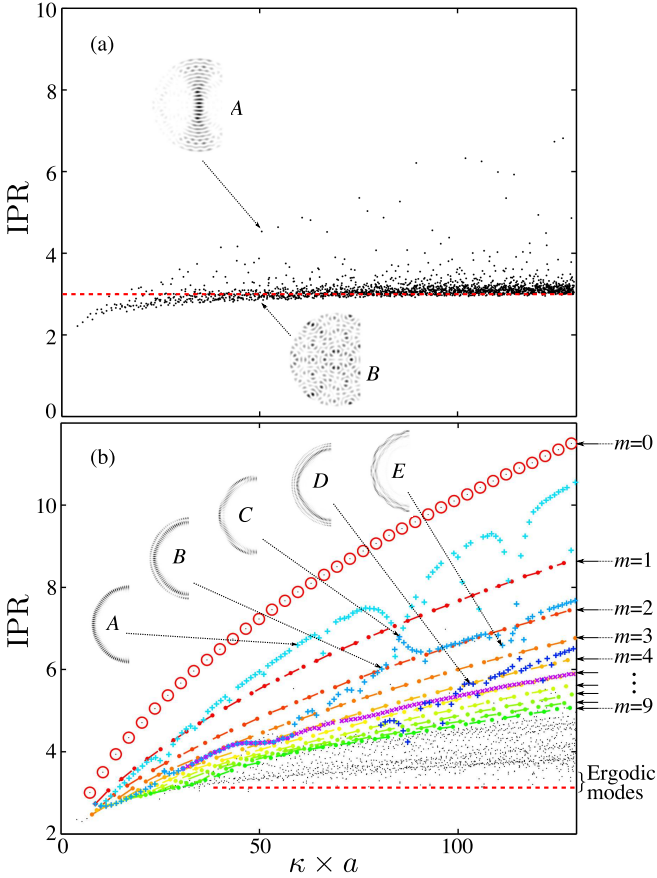}
\caption{Inverse Participation Ratio (IPR) as a function of the transverse wave number $\kappa\times a$. (a) Chaotic case ($\gamma=3/2$) ; As expected, the IPR concentrates in the vicinity of IPR=3 except for the modes built on neutral orbits (upper inset, A). Inset B shows a generic ergodic mode. (b) Mixed case ($\gamma=0.95$) ; Families of modes are distinguishable as shown by the different markers: red circles correspond to 2-PO regular modes, from red to green dots correspond to higher order regular modes denoted from $m=1$ to $m=9$, purple crosses correspond to modes built on the 6-PO, light, medium and dark blue crosses correspond to modes localized on quasi-periodic marginally stable orbits (insets $A$ to $E$).}
\label{IPRtot}
\end{figure}
To investigate the origin of this deviation in the tail of the IPR distribution, we report in Figure \ref{IPRtot} the value of the IPR of each individual mode as a function of $\kappa\times a$ for the chaotic [Fig. \ref{IPRtot}(a)] and mixed [Fig. \ref{IPRtot}(b)] fibers. As expected for the chaotic case, the IPR tends to an asymptotic value of 3 for most of the modes except for a small amount of them. The modes presenting larger values of the IPR [Fig. \ref{IPRtot}(a)] are some specific spatially localized modes built upon quasi-stable trajectories which follow neutral orbits. They are called bouncing ball modes, and take place along the full diameter of the fiber [see inset A in Fig. \ref{IPRtot}(a)]. They thus can be seen as a reminiscence from the regular circular billiard and do not follow the standard gaussian statistics. Figure \ref{IPRtot}(b) presents the IPR as a function of $\kappa\times a$ for the mixed fiber. In a surprising way, one observes a striking structuration of the values of the IPR for individual modes. Moreover, one distinguishes some families of modes following a monotonous behavior. The upper red circles all correspond to the family of regular modes along the 2-PO. The high values of the IPR confirm a strong spatial localization of intensity of these modes, also labelled as the fundamental gaussian modes $u_{0,p}$. The dots in shades of colors from red to green correspond to the IPR calculated for the higher order modes $u_{m,p}$ with $m$ varying from $m=1$ to $m=9$. The regularity of the IPR as a function of $\kappa\times a$ is also observed for higher values of $m$, the maximum being given for $\kappa_{m,0}=k_0\sqrt{\left(n^2_{\rm co}-n^2_{\rm cl}\right)}$, but they are not pointed out in the figure for the sake of clarity. The purple crosses correspond to the modes localized in the vicinity of the 6-bounce PO and follow a well-distinguished behavior as well. The crosses in shades of blue correspond, as pointed out by the inset images labelled from $A$ to $E$, to quasi-periodic marginally stable orbits localized on the boundaries of the cavity. The light blue crosses correspond to modes with a single crown ($A$), as the medium and dark blue crosses correspond respectively to double ($B$) and triple ($D$) crowns. When the crosses intersect, the corresponding modes are a mix of single and double crown modes (resp. double and triple) as shown for the mode $C$ (resp. $E$).
It is important to note that this behavior of strong deviation from the standard value IPR=3 is commonly encountered is systems presenting a strong enough \textsl{disorder} to promote the existence of (Anderson-) localized modes. Here  the \textsl{order}, by means of the regular modes, is surprisingly responsible for the deviation of the IPR.
The modes corresponding to the lower IPRs -- marked in black, right above IPR=3 (red dashed line) -- correspond to chaotic modes that present an ergodic behavior. The density of points around IPR=3 is much less important than above, which confirms that ergodic modes are a minority in a mixed system characterized by the parameter $\gamma=0.95$.

\begin{figure}
\centering
\includegraphics[width=\columnwidth]{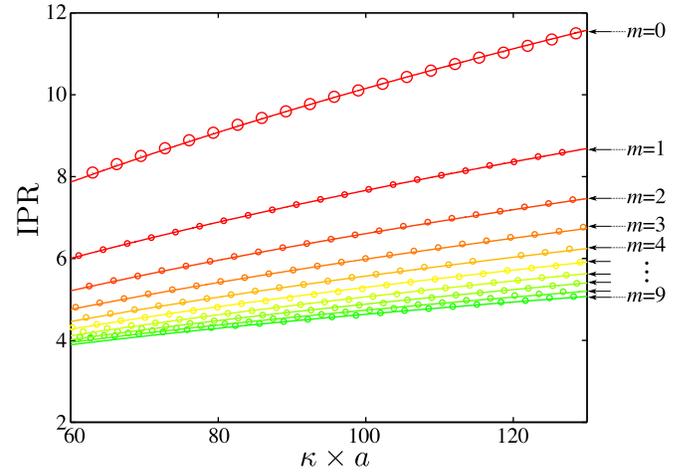}
\caption{IPR as a function of $\kappa\times a$ for the regular modes and for high values of $\kappa$. The dots and circles correspond to the points presented in Fig. \ref{IPRtot}(b) and the continuous lines correspond to the expression (\ref{IPR_HO_modes}).}
\label{IPR_RegFit}
\end{figure}

By use of a heuristic model, we derive an analytic expression for the value of IPR for the high-$\kappa$ regular modes along 2-PO with respect to $\kappa_{0,p}$. We assume that the modes are spatially localized on a surface $\mathcal{S}_{\rm loc}$. Thus, the intensity of each mode equals $I_0$ on $\mathcal{S}_{\rm loc}$, and $0$ elsewhere. Then, $I_0=\mathcal{S}/\mathcal{S}_{\rm loc}$ and using equation (\ref{eqIPR}), one gets:
\begin{equation}
{\rm IPR}=\frac{\mathcal{S}}{\mathcal{S}_{\rm loc}}
\end{equation}
The analogy of these modes with the fundamental gaussian modes $u_{0,p}$ of the stable resonator (see section \ref{Resonator}), allows one to derive an analytic expression for $\mathcal{S}_{\rm loc}$:
\begin{eqnarray}
\mathcal{S}_{\rm loc}&=&2\int_0^{d}w(x)dx\nonumber\\
&=&2w_0x_{\rm r}\left(\frac{d}{x_{\rm r}}\sqrt{1+\left(\frac{d}{x_{\rm r}}\right)^2}+\mathrm{arcsinh}\left(\frac{d}{x_{\rm r}}\right)\right)\nonumber\\
&=&w_0L_{\rm loc}
\end{eqnarray}
where $L_{\rm loc}$ is an effective ``localization length''. Finally, one gets:
\begin{equation}
{\rm IPR}=\frac{\mathcal{S}\sqrt{\kappa}}{L_{\rm loc}\sqrt{2x_{\rm r}}}\label{IPR}
\end{equation}
We obtain a perfect agreement between this expression and the IPR of the 2-PO regular modes $u_{m=0,p}$ [large red circles and red continuous line in Fig. \ref{IPR_RegFit}].\\
Moreover, for large values of $\kappa$, the IPR appears to structure itself by presenting a regular evolution for each family of modes. The IPR for the modes of a given $m$ value are following a monotonous curve which is perfectly adjusted by an empirical expression:
\begin{equation}
{\rm IPR}(u_{m,p})=c_{m}\kappa^{\xi_{m}}
\label{IPR_HO_modes}
\end{equation}
with $\xi_{m}=-0.017\times m+\frac{1}{2}$ and $c_m$ a constant.

In Figure \ref{IPR_RegFit}, the red circles correspond to the 2-PO regular modes $u_{0,p}$ and the line corresponds to the plot of the expression (\ref{IPR}). The dots in shades of colors from red to green are the IPRs of higher order modes as shown in Fig. \ref{IPRtot} and the lines are given by plotting the expression (\ref{IPR_HO_modes}). As one can note, the agreement between the values of the IPR and the analytic curves is very good.


\section{Experimental observation and manipulation of the regular modes of the mixed fiber}
\label{Experimental}

In order to study experimentally the light propagation into a mixed fiber, we manufactured a multimode optical fiber whose transverse shape is a truncated circle with $\gamma=0.95$. The illumination is made with a 2mW HeNe cw laser ($\lambda_0=633$nm). The polarization is controlled through a polarizer and a $\lambda/2$ wave-plate. The laser beam, of initial diameter 2mm, is extended in order to use all the surface of a Spatial Light Modulator (SLM, Amplitude only). The SLM is used to shape the intensity profile of the illumination beam. The beam is focused into the fiber thanks to a microscope objective in order to fit the surface of the fiber's core. The angle of the initial beam is controlled by the angle of the input end of the fiber with respect to the optical axis. It allows to fix the average transverse wavenumber of the illumination and thus to select the range of excited modes. A spatial modulation can be imposed through the SLM in order to shape the spatial distribution of the field intensity and then select a family of excited modes.
At the output, the beam is collected by a microscope objective and sent to a CCD Camera. The NF is then directly collected. A supplementary lens is needed in order to collect the FF.

\begin{figure}
\centering
\includegraphics[width=\columnwidth]{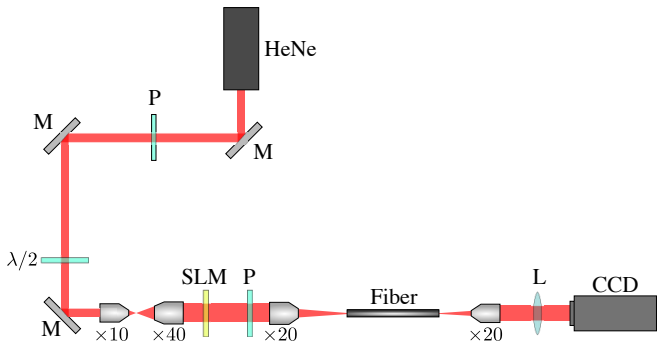}
\caption{Experimental setup. P: polarizer; M: mirror; $\lambda/2$: half-wavelength waveplate; $\times 10$, $\times 20$, $\times 40$: microscope objectives; SLM: Spatial Light Modulator; L: lens needed to collect the FF; CCD: CCD camera.}
\label{SchemaManip}
\end{figure}

\begin{figure*}
\centering
\includegraphics[width=\textwidth]{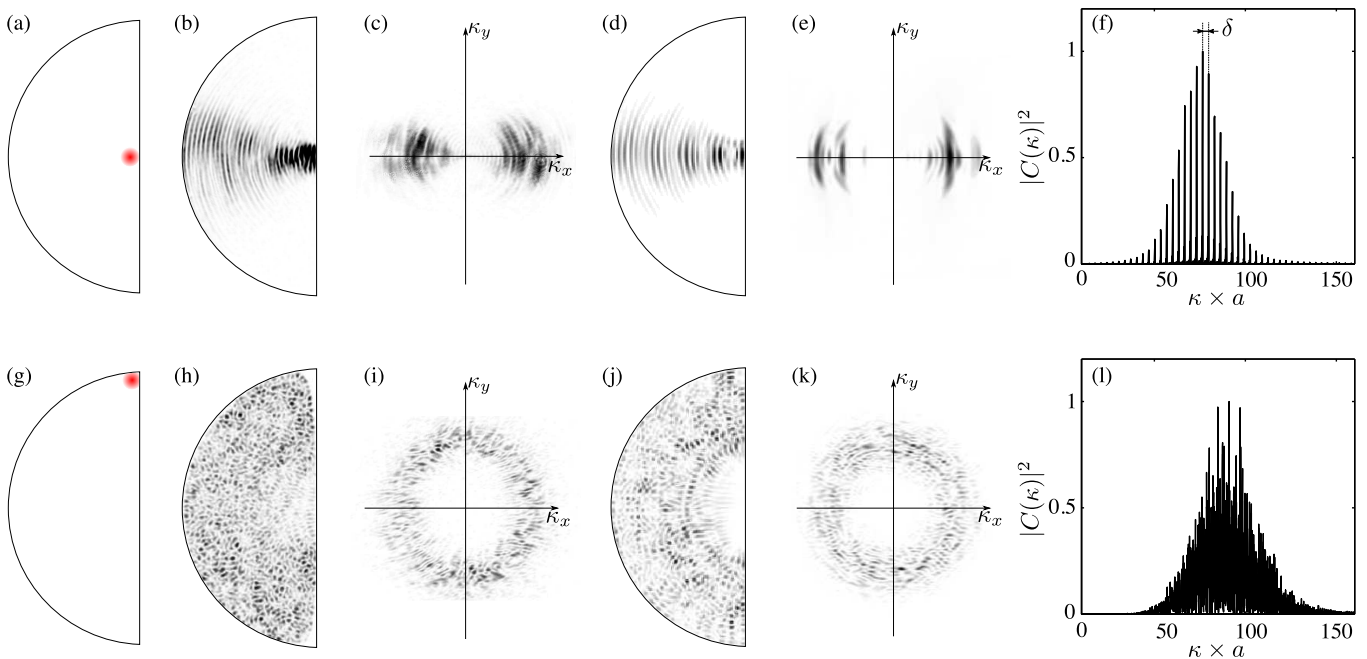}
\caption{Initial illumination beams, NF and FF observed at the output of the fiber, corresponding numerical simulations and numerical calculated spectra, for an illumination favorable to the excitation of a family of (a-f) regular modes and (g-l) ergodic modes.}
\label{Experiment}
\end{figure*}

At first, we did not use the SLM, and we only control the illumination by the position of a focused beam in front of the input end of the fiber. Figures \ref{Experiment}(a) and (g) show two typical excitations. The gaussian shape is obtained at the focal point of a microscope objective, and a tilt is given in order to select the range of transverse wave vectors propagating into the fiber. 
Figure \ref{Experiment}(b) presents the NF measured at the fiber's output for the first illumination [Fig. \ref{Experiment}(a)]. The spatial repartition of the NF is characteristic of a superposition of regular modes of the 2-PO. One can see the large amount of intensity near the straight boundary which is a signature of the excitation of these modes. Indeed, all the regular modes along the 2-PO present such a high intensity in the vicinity of the straight boundary, along the $x$-axis. Moreover, one can note the circular arc-like pattern also present on Figures \ref{modes}(a) and (e). Figure \ref{Experiment}(c) presents the FF intensity pattern of the outgoing wave. The anisotropic configuration along the $\kappa_x$-axis confirms that the field builds up on the 2-PO. The spatial extension of the FF testifies that numerous gaussian-like modes of the type $u_{0,p}(x,y)$ have been excited. Figures \ref{Experiment}(d) and (e) show the NF and FF resulting from numerical simulations. Using a standard beam propagation method algorithm \cite{Feit1978}, we simulate the propagation of an initial gaussian beam along the optical fiber. The parameters here are given in order to reproduce the experimental initial condition. As one can see, the agreement between the numerical and experimental results is very good, without any adjustable parameter. Figure \ref{Experiment}(f) presents the numerical spectrum calculated through the standard method \cite{Feit1980}. More precisely, $|\mathcal{C}(\kappa)|^2$ is the square modulus of the Fourier transform of the correlation function defined as the overlap between the outgoing field and the initial condition. The observed Fabry-Perot-like spectrum confirms that the analogy between the 2-PO regular modes of the mixed fiber and the gaussian modes of a stable optical cavity is relevant and one can measure a free spectral range $\delta=\left(\Delta\kappa\right)\times a=\pi a/d=\pi/\gamma$ where $\Delta\kappa=\kappa_{m,p+1}-\kappa_{m,p}$ as defined in equation (\ref{kt}).

Figure \ref{Experiment}(h) presents the NF collected for the second illumination [Fig. \ref{Experiment}(g)]. Here, the pattern is completely different and the spatial distribution of intensity is statistically uniform, made of grains of light of random amplitude and size. This speckle-like pattern results from the superposition of ergodic modes [see Fig. \ref{Modes}(j)]. The FF [Fig. \ref{Experiment}(i)] confirms this assumption, by presenting an isotropic distribution of wavevectors whose modulus are radially confined around an average wavenumber $\bar \kappa$. Figures \ref{Experiment}(j) and (k) present the associated numerical simulations and show a good agreement with the experimental results. The numerical spectrum [Figure \ref{Experiment}(l)] does not present any predominant structuration. Moreover, it shows no discrimination of the modes: all the modes in the range of $\kappa$ corresponding to the illumination are excited. The three regularly-spaced peaks one can see around $\kappa\times a=90$ are a reminiscence of a periodic orbit that have been excited and which is responsible for the caustic visible on Figure \ref{Experiment}(j).

The evaluation of the IPR from experiments for a superposition of modes is complicated due to the beating between the excited modes along the propagation. However, it is always possible to get an estimation of the IPR at the fiber's output and to compare the result with the numerical simulations for given propagation lengths. We note that the IPR is always around 3 for the ergodic-like spatial repartition of the intensity and widely above for the superposition of regular modes of the 2-PO. This result confirms the very different nature of these two kinds of behaviors.\\

From the spectra, we aim to extract some information on the underlying geometrical signature of the modes. The spectra presented in Figures \ref{Experiment}(f) and (l) are, as a matter of fact, equivalent to the density of states $n(\kappa)$ so as $n(\kappa)=|\mathcal{C}(\kappa)|^2$. An other expression of this density of states reads \cite{Gutzwiller1990} :
\begin{equation}
	n(\kappa)=n_0(\kappa)+\sum_j \ell_j\mathcal{A}_j\exp{\left( i\kappa \ell_j \right)}
	\label{DOS}
\end{equation}
where the sum is running over the periodic orbits. $\ell_j$ is the total transverse optical length of orbit $j$ and $\mathcal{A}_j$ encompasses a classical amplitude related to the stability of the orbit and a phase associated to caustics and reflections. The first part of the density, $n_0(\kappa)$ is given by the so-called Weyl formula \cite{Balian1971} :
\begin{equation}
	n_0(\kappa)=\frac{\mathcal{S}}{2\pi}\kappa-\frac{\mathcal{P}}{4\pi}
	\label{Weyl}
\end{equation}
where $\mathcal{P}$ is the perimeter of the cross section of the fiber.
As one can see in equation (\ref{DOS}), a relevant way to characterize the presence or not of any regularity in the underlying ray dynamics is to calculate the ``length spectra'', as the Fourier transform of $n(\kappa)$ in order to get the quantity $\mathcal{L}(\ell)$ as a function of $\ell$. $\mathcal{L}(\ell)$ will display peaks at the corresponding orbit length. Thus, the geometrical length of the periodic orbits can be directly extracted from the measure of the spectrum $n(\kappa)$ which is calculated independently from the knowledge of the underlying periodic orbits \cite{Lebental2007a}.

\begin{figure}
\centering
\includegraphics[width=\columnwidth]{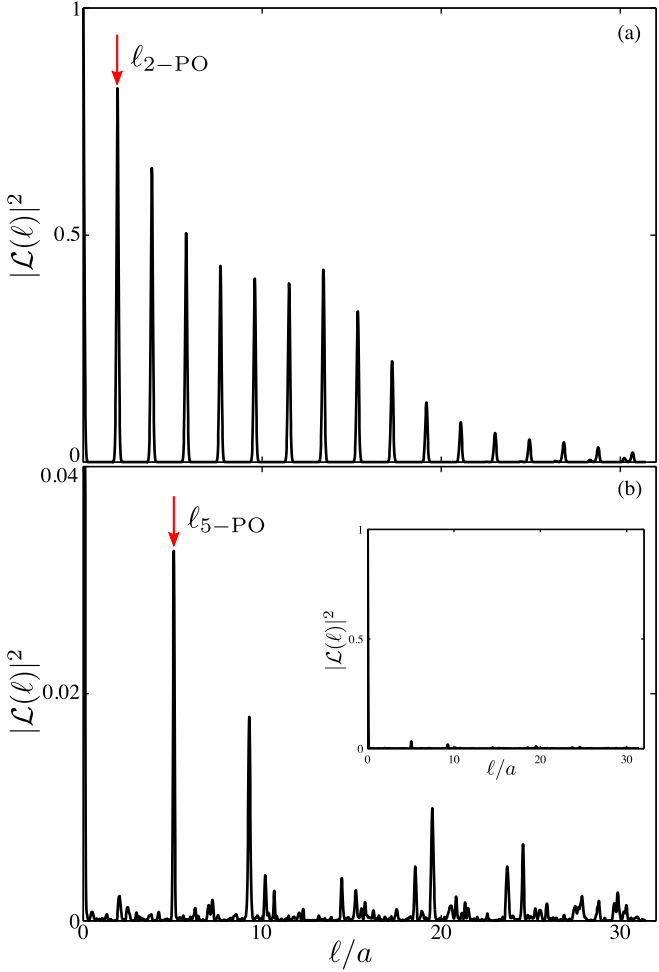}
\caption{Length spectra corresponding to spectra presented in (a) Fig. \ref{Experiment}(f): one can see the regularly distributed peaks corresponding for the first to the length of the 2-PO and for the others to the harmonics ; and in (b) Fig. \ref{Experiment}(l): one can see that no specific length exists (see inset) even if a zoom shows some peaks distributed in an arbitrary way. The arrows point the 2-PO and 5-PO respectively.}
\label{LengthSpectra}
\end{figure}
Figure \ref{LengthSpectra}(a) presents the length spectrum calculated from Fig. \ref{Experiment}(f). The first peak corresponds to the length of the 2-PO (pointed out by the red dashed line), that is $\ell/a=2\times \gamma=1.9$ (the factor $2$ appearing for the round trip) and the other peaks being the harmonics. It is thus obvious that the only contribution to the spatial distribution of the field is due to a superposition of regular modes of the 2-PO. Figure \ref{LengthSpectra}(b) shows the length spectrum associated to the spectrum of Fig. \ref{Experiment}(l). Some peaks still appear as residual resonances not predominant in the dynamics. The length spectrum with the same scale as the latter is shown in the inset and it is clear that no specific trajectories are promoted.

\begin{figure}
\centering
\includegraphics[width=\columnwidth]{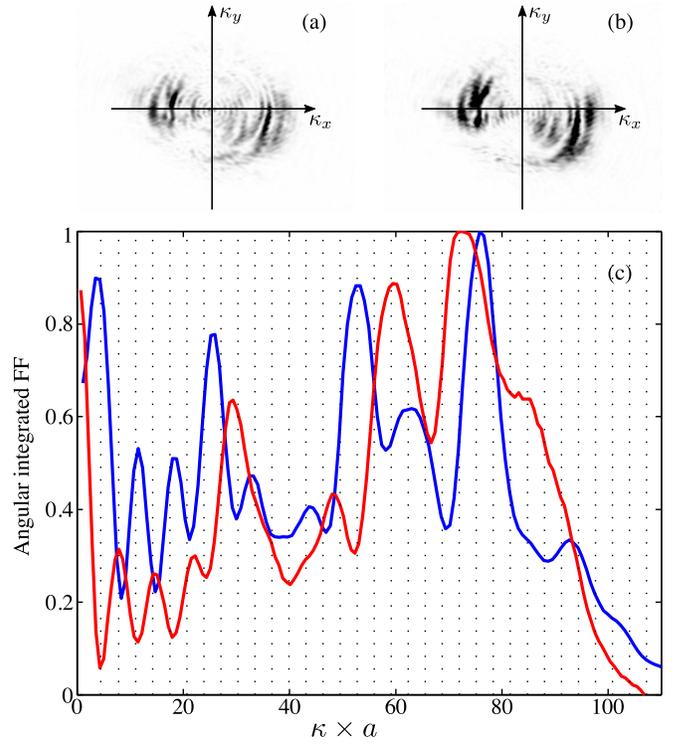}
\caption{(a,b) Experimental FF measured for two different spatial modulations of the illumination beam. (c) Angular integrated FF. Blue line corresponds to case (a) and red line to case (b).}
\label{AngIntExp}
\end{figure}

As shown right above, a beam focused close to the truncation -- which corresponds to the location of the gaussian transverse modes waist -- leads to the excitation of 2-PO regular modes. The width of the spectrum (the number of excited modes) then depends on the size of the illumination beam. A fine tuning of the illumination is obtained by adding a spatial modulation along the $x$-axis by use of a SLM: it allows one to select the order of the excited modes. Figures \ref{AngIntExp}(a and b) show examples of the FF for two different modulations associated to two different transverse wavenumbers $\kappa$. We notice that the mean radius of the intensity repartition in the far-field, that is the mean value of $\kappa$, is slightly different. 
A low-resolution spatial frequency spectrum is obtained from the angular integration of the far-field [Fig. \ref{AngIntExp}(c)]. It allows to get rough informations on the transverse wavenumbers characterizing the propagated field. The blue line (resp. red line) corresponds to the FF presented in Fig. \ref{AngIntExp}(a) (resp. (b)).

In Fig. \ref{AngIntExp}, modulations clearly appear both in the $\kappa_x$-direction in the FFs and in their angular integrations. It points out that only some specific modes (the 2-PO regular modes) have been selected. By superimposing the expected values of $\kappa$ for the 2-PO regular modes calculated for our experimental system using eq. (\ref{kt}), one can notice that most of the maxima in the angular integrated far-field correspond to some $\kappa_{0,p}$ values associated to the $u_{0,p}(x,y)$ 2-PO regular modes. As confirmed by the 2D representations of the FF [Figs. \ref{AngIntExp}(a,b)], 2-PO regular modes are essentially excited. 
By changing the modulation, we slightly change the location of the maximum of intensity of the illumination beam and then excite preferentially regular modes with inverse parity.

\begin{figure}
\centering
\includegraphics[width=\columnwidth]{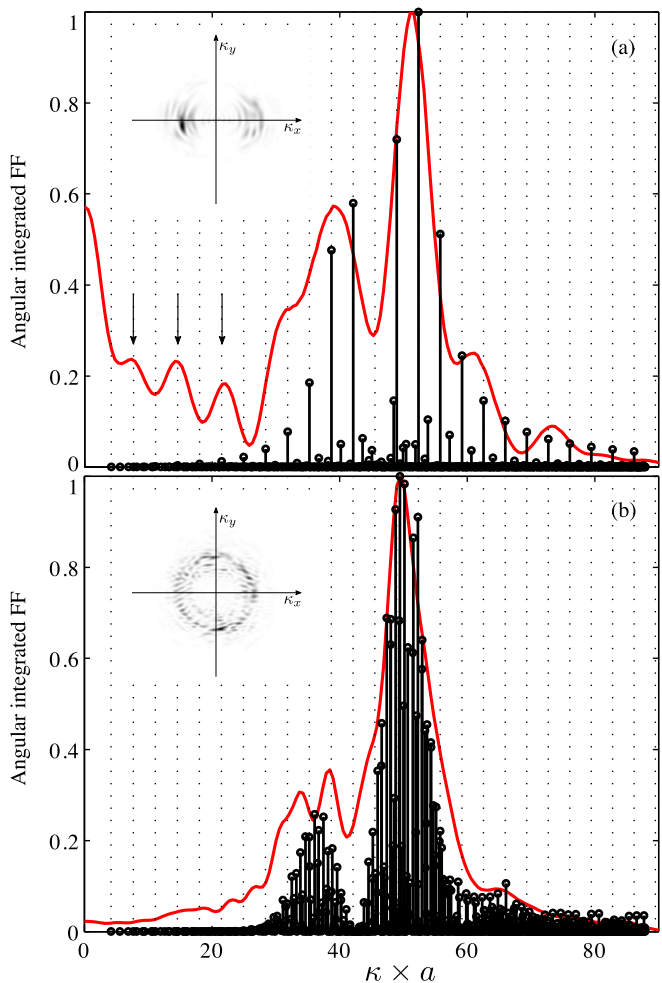}
\caption{Angular integrated FF obtained by numerical simumations. (a) For an illumination favorable to the excitation of regular modes. Inset shows the FF ; (b) For a modulation along the $y$-direction, so that no regular mode is being excited. Inset shows the FF.}
\label{AngIntNum}
\end{figure}

We perform a numerical simulation to analyze the experimental angular far-field integration. We project two initial conditions analogous to the ones used in the experiment -- that is an asymetric gaussian beam stretched and modulated in the $x$-direction -- on the calculated modes to get the weights $c_n$ of each mode in the propagating field $\psi(x,y,z)$ describes as :
\begin{equation}
\psi(x,y,z)=\sum_nc_n\phi_n(x,y)\exp{\left(-j\beta_nz\right)}\label{propfield}
\end{equation}
where $\phi_n$ is a guided mode and $\beta_n=\sqrt{k_0^2-\kappa_n^2}$ its corresponding constant of propagation.
Numerically, we can superimpose the angular integrated far-field with the calculated weights $c_n$ of each modes. Figure \ref{AngIntNum}(a) shows a behavior analogous to the experimental angular integrated far-fields. The peaks observed for low value of $\kappa\times a$ are also present and pointed by the vertical arrows. Actually, they also appear in the angular integrated far-field calculated for a unique 2-PO regular mode and are a consequence of the gaussian beam nature of the 2-PO regular modes along the stable trajectory. 
To compare, scar modes, which are the result of constructive interferences between two counter-propagative plane waves, are characterized by two symmetric peaks in the FF. Consequently, a unique peak would appear in the angular integrated FF.
In Fig. \ref{AngIntNum}, we observe a broaden peak in the integrated FF, which is due to the interferences between the excited modes preferentially being 2-PO regular modes.
Finally, we report in Fig. \ref{AngIntNum}(b), the angular integrated far-field for an initial modulated beam out of the 2-PO direction (precisely, in the upper part of the fiber, with a modulation having the same $\kappa$ value but along the $y$-direction to avoid the 2-PO). We observe the vanishment of the first peaks associated to the characteristic signature of the 2-PO regular modes and an enlarged peak that corresponds to a superposition of a great number of arbitrary excited modes around the main $\kappa$ value.


\section{Conclusion and discussion}
\label{Conclusion}

In this paper we have presented numerical and experimental investigations of the modes of a highly multimode fiber whose transverse cross section is designed to induced a complex ray dynamics. This multimode optical fiber is clearly an ideal experimental support for this kind of study, as it allows an easy-to-implement setup, with a controllable excitation and a direct visualization. Moreover, this system is versatile in the sense that a slight change in the level of the truncation leads to the exploration of different types of dynamics. One thus has access, with the same experimental system, to mixed, regular as well as chaotic dynamics. Here, we focused our work on the study of a mixed dynamics. In particular, we concentrated the analysis on a family of regular modes of the system, that presents a spatially localized intensity pattern, as well as a Fabry-Perot-like discrete spectrum. An analogy with the gaussian modes of an optical resonator leads to the derivation of an analytical expression for the regular modes as well as a resonant condition justifying the regular distributed spectrum. The strong spatial localization of the regular modes finds some signatures in the study of the Inverse Participation Ratio, by largely exceeding the value predicted by the Random Matrix Theory. By coupling these results with the analogy with the gaussian modes, we then were able to give an expression for the IPR as a function of the wavenumber for the regular modes.

With a suitable shaping of the initial beam, we experimentally demonstrated that these modes can be selectively excited, and that they are robust to mode coupling. This ease of shielding the other modes open the way to multiple applications in optical telecommunications. For instance, the regular modes of the mixed optical fiber are perfectly suitable for Mode Division Multiplexing (MDM) and would allow the use of much more transmission channels than already achieved in conventional devices. Moreover, these modes can benefit from a selective optical amplification, by optimizing the spatial overlap with a gain medium. To do so, one simply has to locate the active medium in the vicinity of the truncation, where the regular modes have their maximum of intensity \cite{Michel2012}. 
In a more fundamental point of view, a nonlinear mixed optical fiber would allow an enhancement of the phenomenon of optical thermalization and condensation of classical waves, as the spatial overlap between the modes is an important parameter \cite{Aschieri2011}.\\

The authors thank gratefully M. Ud\'e and S. Trzesien for their involvement in the preform and fiber manufacture.

\bibliographystyle{apsrev4-1}
\bibliography{library}

\begin{thebibliography}{44}%
\makeatletter
\providecommand \@ifxundefined [1]{%
 \@ifx{#1\undefined}
}%
\providecommand \@ifnum [1]{%
 \ifnum #1\expandafter \@firstoftwo
 \else \expandafter \@secondoftwo
 \fi
}%
\providecommand \@ifx [1]{%
 \ifx #1\expandafter \@firstoftwo
 \else \expandafter \@secondoftwo
 \fi
}%
\providecommand \natexlab [1]{#1}%
\providecommand \enquote  [1]{``#1''}%
\providecommand \bibnamefont  [1]{#1}%
\providecommand \bibfnamefont [1]{#1}%
\providecommand \citenamefont [1]{#1}%
\providecommand \href@noop [0]{\@secondoftwo}%
\providecommand \href [0]{\begingroup \@sanitize@url \@href}%
\providecommand \@href[1]{\@@startlink{#1}\@@href}%
\providecommand \@@href[1]{\endgroup#1\@@endlink}%
\providecommand \@sanitize@url [0]{\catcode `\\12\catcode `\$12\catcode
  `\&12\catcode `\#12\catcode `\^12\catcode `\_12\catcode `\%12\relax}%
\providecommand \@@startlink[1]{}%
\providecommand \@@endlink[0]{}%
\providecommand \url  [0]{\begingroup\@sanitize@url \@url }%
\providecommand \@url [1]{\endgroup\@href {#1}{\urlprefix }}%
\providecommand \urlprefix  [0]{URL }%
\providecommand \Eprint [0]{\href }%
\providecommand \doibase [0]{http://dx.doi.org/}%
\providecommand \selectlanguage [0]{\@gobble}%
\providecommand \bibinfo  [0]{\@secondoftwo}%
\providecommand \bibfield  [0]{\@secondoftwo}%
\providecommand \translation [1]{[#1]}%
\providecommand \BibitemOpen [0]{}%
\providecommand \bibitemStop [0]{}%
\providecommand \bibitemNoStop [0]{.\EOS\space}%
\providecommand \EOS [0]{\spacefactor3000\relax}%
\providecommand \BibitemShut  [1]{\csname bibitem#1\endcsname}%
\let\auto@bib@innerbib\@empty
\bibitem [{\citenamefont {Gros}\ \emph {et~al.}(2014)\citenamefont {Gros},
  \citenamefont {Legrand}, \citenamefont {Mortessagne}, \citenamefont
  {Richalot},\ and\ \citenamefont {Selemani}}]{Gros2014}%
  \BibitemOpen
  \bibfield  {author} {\bibinfo {author} {\bibfnamefont {J.~B.}\ \bibnamefont
  {Gros}}, \bibinfo {author} {\bibfnamefont {O.}~\bibnamefont {Legrand}},
  \bibinfo {author} {\bibfnamefont {F.}~\bibnamefont {Mortessagne}}, \bibinfo
  {author} {\bibfnamefont {E.}~\bibnamefont {Richalot}}, \ and\ \bibinfo
  {author} {\bibfnamefont {K.}~\bibnamefont {Selemani}},\ }\href@noop {}
  {\bibfield  {journal} {\bibinfo  {journal} {Wave Motion}\ }\textbf {\bibinfo
  {volume} {51}},\ \bibinfo {pages} {664} (\bibinfo {year} {2014})}\BibitemShut
  {NoStop}%
\bibitem [{\citenamefont {Selemani}\ \emph {et~al.}(2015)\citenamefont
  {Selemani}, \citenamefont {Gros}, \citenamefont {Richalot}, \citenamefont
  {Legrand}, \citenamefont {Picon},\ and\ \citenamefont
  {Mortessagne}}]{Selemani2015}%
  \BibitemOpen
  \bibfield  {author} {\bibinfo {author} {\bibfnamefont {K.}~\bibnamefont
  {Selemani}}, \bibinfo {author} {\bibfnamefont {J.-B.}\ \bibnamefont {Gros}},
  \bibinfo {author} {\bibfnamefont {E.}~\bibnamefont {Richalot}}, \bibinfo
  {author} {\bibfnamefont {O.}~\bibnamefont {Legrand}}, \bibinfo {author}
  {\bibfnamefont {O.}~\bibnamefont {Picon}}, \ and\ \bibinfo {author}
  {\bibfnamefont {F.}~\bibnamefont {Mortessagne}},\ }\href@noop {} {\bibfield
  {journal} {\bibinfo  {journal} {IEEE Trans. Electromagn. Compat.}\ }\textbf
  {\bibinfo {volume} {57}},\ \bibinfo {pages} {3} (\bibinfo {year}
  {2015})}\BibitemShut {NoStop}%
\bibitem [{\citenamefont {Cao}\ and\ \citenamefont {Wiersig}(2015)}]{Cao2015}%
  \BibitemOpen
  \bibfield  {author} {\bibinfo {author} {\bibfnamefont {H.}~\bibnamefont
  {Cao}}\ and\ \bibinfo {author} {\bibfnamefont {J.}~\bibnamefont {Wiersig}},\
  }\href@noop {} {\bibfield  {journal} {\bibinfo  {journal} {RMP}\ }\textbf
  {\bibinfo {volume} {87}},\ \bibinfo {pages} {61} (\bibinfo {year}
  {2015})}\BibitemShut {NoStop}%
\bibitem [{\citenamefont {Gmachl}\ \emph {et~al.}(1998)\citenamefont {Gmachl},
  \citenamefont {Capasso}, \citenamefont {Narimanov}, \citenamefont {Noeckel},
  \citenamefont {Stone}, \citenamefont {Faist}, \citenamefont {Sivco},\ and\
  \citenamefont {Cho}}]{Gmachl1998}%
  \BibitemOpen
  \bibfield  {author} {\bibinfo {author} {\bibfnamefont {C.}~\bibnamefont
  {Gmachl}}, \bibinfo {author} {\bibfnamefont {F.}~\bibnamefont {Capasso}},
  \bibinfo {author} {\bibfnamefont {E.~E.}\ \bibnamefont {Narimanov}}, \bibinfo
  {author} {\bibfnamefont {J.~U.}\ \bibnamefont {Noeckel}}, \bibinfo {author}
  {\bibfnamefont {a.~D.}\ \bibnamefont {Stone}}, \bibinfo {author}
  {\bibfnamefont {J.}~\bibnamefont {Faist}}, \bibinfo {author} {\bibfnamefont
  {D.~L.}\ \bibnamefont {Sivco}}, \ and\ \bibinfo {author} {\bibfnamefont
  {A.~Y.}\ \bibnamefont {Cho}},\ }\href@noop {} {\bibfield  {journal} {\bibinfo
   {journal} {Science}\ }\textbf {\bibinfo {volume} {5}},\ \bibinfo {pages}
  {1556} (\bibinfo {year} {1998})}\BibitemShut {NoStop}%
\bibitem [{\citenamefont {Harayama}\ \emph {et~al.}(2003)\citenamefont
  {Harayama}, \citenamefont {Fukushima}, \citenamefont {Davis}, \citenamefont
  {Vaccaro}, \citenamefont {Miyasaka}, \citenamefont {Nishimura},\ and\
  \citenamefont {Aida}}]{Harayama2003b}%
  \BibitemOpen
  \bibfield  {author} {\bibinfo {author} {\bibfnamefont {T.}~\bibnamefont
  {Harayama}}, \bibinfo {author} {\bibfnamefont {T.}~\bibnamefont {Fukushima}},
  \bibinfo {author} {\bibfnamefont {P.}~\bibnamefont {Davis}}, \bibinfo
  {author} {\bibfnamefont {P.~O.}\ \bibnamefont {Vaccaro}}, \bibinfo {author}
  {\bibfnamefont {T.}~\bibnamefont {Miyasaka}}, \bibinfo {author}
  {\bibfnamefont {T.}~\bibnamefont {Nishimura}}, \ and\ \bibinfo {author}
  {\bibfnamefont {T.}~\bibnamefont {Aida}},\ }\href@noop {} {\bibfield
  {journal} {\bibinfo  {journal} {PRE}\ }\textbf {\bibinfo {volume} {67}},\
  \bibinfo {pages} {015207} (\bibinfo {year} {2003})}\BibitemShut {NoStop}%
\bibitem [{\citenamefont {Ge}\ \emph {et~al.}(2015)\citenamefont {Ge},
  \citenamefont {Sarma},\ and\ \citenamefont {Cao}}]{Ge2015}%
  \BibitemOpen
  \bibfield  {author} {\bibinfo {author} {\bibfnamefont {L.}~\bibnamefont
  {Ge}}, \bibinfo {author} {\bibfnamefont {R.}~\bibnamefont {Sarma}}, \ and\
  \bibinfo {author} {\bibfnamefont {H.}~\bibnamefont {Cao}},\ }\href@noop {}
  {\bibfield  {journal} {\bibinfo  {journal} {Optica}\ }\textbf {\bibinfo
  {volume} {2}},\ \bibinfo {pages} {323} (\bibinfo {year} {2015})}\BibitemShut
  {NoStop}%
\bibitem [{\citenamefont {Doya}\ \emph
  {et~al.}(2001{\natexlab{a}})\citenamefont {Doya}, \citenamefont {Legrand},\
  and\ \citenamefont {Mortessagne}}]{Doya2001}%
  \BibitemOpen
  \bibfield  {author} {\bibinfo {author} {\bibfnamefont {V.}~\bibnamefont
  {Doya}}, \bibinfo {author} {\bibfnamefont {O.}~\bibnamefont {Legrand}}, \
  and\ \bibinfo {author} {\bibfnamefont {F.}~\bibnamefont {Mortessagne}},\
  }\href@noop {} {\bibfield  {journal} {\bibinfo  {journal} {Opt. Lett.}\
  }\textbf {\bibinfo {volume} {26}},\ \bibinfo {pages} {872} (\bibinfo {year}
  {2001}{\natexlab{a}})}\BibitemShut {NoStop}%
\bibitem [{\citenamefont {B\"{a}cker}\ \emph {et~al.}(2009)\citenamefont
  {B\"{a}cker}, \citenamefont {Ketzmerick}, \citenamefont {L\"{o}ck},
  \citenamefont {Wiersig},\ and\ \citenamefont {Hentschel}}]{Backer2009}%
  \BibitemOpen
  \bibfield  {author} {\bibinfo {author} {\bibfnamefont {A.}~\bibnamefont
  {B\"{a}cker}}, \bibinfo {author} {\bibfnamefont {R.}~\bibnamefont
  {Ketzmerick}}, \bibinfo {author} {\bibfnamefont {S.}~\bibnamefont
  {L\"{o}ck}}, \bibinfo {author} {\bibfnamefont {J.}~\bibnamefont {Wiersig}}, \
  and\ \bibinfo {author} {\bibfnamefont {M.}~\bibnamefont {Hentschel}},\
  }\href@noop {} {\bibfield  {journal} {\bibinfo  {journal} {PRA}\ }\textbf
  {\bibinfo {volume} {79}},\ \bibinfo {pages} {063804} (\bibinfo {year}
  {2009})}\BibitemShut {NoStop}%
\bibitem [{\citenamefont {Aschi\'{e}ri}\ and\ \citenamefont
  {Doya}(2013)}]{Aschieri2013}%
  \BibitemOpen
  \bibfield  {author} {\bibinfo {author} {\bibfnamefont {P.}~\bibnamefont
  {Aschi\'{e}ri}}\ and\ \bibinfo {author} {\bibfnamefont {V.}~\bibnamefont
  {Doya}},\ }\href@noop {} {\bibfield  {journal} {\bibinfo  {journal} {JOSAB}\
  }\textbf {\bibinfo {volume} {30}},\ \bibinfo {pages} {3161} (\bibinfo {year}
  {2013})}\BibitemShut {NoStop}%
\bibitem [{\citenamefont {Song}\ \emph {et~al.}(2014)\citenamefont {Song},
  \citenamefont {Liu}, \citenamefont {Gu}, \citenamefont {Zhang},\ and\
  \citenamefont {Xiao}}]{Song2014}%
  \BibitemOpen
  \bibfield  {author} {\bibinfo {author} {\bibfnamefont {Q.}~\bibnamefont
  {Song}}, \bibinfo {author} {\bibfnamefont {S.}~\bibnamefont {Liu}}, \bibinfo
  {author} {\bibfnamefont {Z.}~\bibnamefont {Gu}}, \bibinfo {author}
  {\bibfnamefont {N.}~\bibnamefont {Zhang}}, \ and\ \bibinfo {author}
  {\bibfnamefont {S.}~\bibnamefont {Xiao}},\ }\href@noop {} {\bibfield
  {journal} {\bibinfo  {journal} {Opt. Lett.}\ }\textbf {\bibinfo {volume}
  {39}},\ \bibinfo {pages} {1149} (\bibinfo {year} {2014})}\BibitemShut
  {NoStop}%
\bibitem [{\citenamefont {Doya}\ \emph {et~al.}(2002)\citenamefont {Doya},
  \citenamefont {Legrand}, \citenamefont {Mortessagne},\ and\ \citenamefont
  {Miniatura}}]{Doya2002}%
  \BibitemOpen
  \bibfield  {author} {\bibinfo {author} {\bibfnamefont {V.}~\bibnamefont
  {Doya}}, \bibinfo {author} {\bibfnamefont {O.}~\bibnamefont {Legrand}},
  \bibinfo {author} {\bibfnamefont {F.}~\bibnamefont {Mortessagne}}, \ and\
  \bibinfo {author} {\bibfnamefont {C.}~\bibnamefont {Miniatura}},\ }\href@noop
  {} {\bibfield  {journal} {\bibinfo  {journal} {PRE}\ }\textbf {\bibinfo
  {volume} {65}},\ \bibinfo {pages} {056223} (\bibinfo {year}
  {2002})}\BibitemShut {NoStop}%
\bibitem [{\citenamefont {Doya}\ \emph
  {et~al.}(2001{\natexlab{b}})\citenamefont {Doya}, \citenamefont {Legrand},
  \citenamefont {Mortessagne},\ and\ \citenamefont {Miniatura}}]{Doya2001a}%
  \BibitemOpen
  \bibfield  {author} {\bibinfo {author} {\bibfnamefont {V.}~\bibnamefont
  {Doya}}, \bibinfo {author} {\bibfnamefont {O.}~\bibnamefont {Legrand}},
  \bibinfo {author} {\bibfnamefont {F.}~\bibnamefont {Mortessagne}}, \ and\
  \bibinfo {author} {\bibfnamefont {C.}~\bibnamefont {Miniatura}},\ }\href@noop
  {} {\bibfield  {journal} {\bibinfo  {journal} {PRL}\ }\textbf {\bibinfo
  {volume} {88}},\ \bibinfo {pages} {014102} (\bibinfo {year}
  {2001}{\natexlab{b}})}\BibitemShut {NoStop}%
\bibitem [{\citenamefont {Michel}\ \emph {et~al.}(2009)\citenamefont {Michel},
  \citenamefont {Doya}, \citenamefont {Tascu}, \citenamefont {Blanc},
  \citenamefont {Legrand},\ and\ \citenamefont {Mortessagne}}]{Michel2009a}%
  \BibitemOpen
  \bibfield  {author} {\bibinfo {author} {\bibfnamefont {C.}~\bibnamefont
  {Michel}}, \bibinfo {author} {\bibfnamefont {V.}~\bibnamefont {Doya}},
  \bibinfo {author} {\bibfnamefont {S.}~\bibnamefont {Tascu}}, \bibinfo
  {author} {\bibfnamefont {W.}~\bibnamefont {Blanc}}, \bibinfo {author}
  {\bibfnamefont {O.}~\bibnamefont {Legrand}}, \ and\ \bibinfo {author}
  {\bibfnamefont {F.}~\bibnamefont {Mortessagne}},\ }\href@noop {} {\bibfield
  {journal} {\bibinfo  {journal} {Appl. Opt.}\ }\textbf {\bibinfo {volume}
  {48}},\ \bibinfo {pages} {G163} (\bibinfo {year} {2009})}\BibitemShut
  {NoStop}%
\bibitem [{\citenamefont {Michel}\ \emph {et~al.}(2012)\citenamefont {Michel},
  \citenamefont {Tascu}, \citenamefont {Doya}, \citenamefont {Aschi\'{e}ri},
  \citenamefont {Blanc}, \citenamefont {Legrand},\ and\ \citenamefont
  {Mortessagne}}]{Michel2012}%
  \BibitemOpen
  \bibfield  {author} {\bibinfo {author} {\bibfnamefont {C.}~\bibnamefont
  {Michel}}, \bibinfo {author} {\bibfnamefont {S.}~\bibnamefont {Tascu}},
  \bibinfo {author} {\bibfnamefont {V.}~\bibnamefont {Doya}}, \bibinfo {author}
  {\bibfnamefont {P.}~\bibnamefont {Aschi\'{e}ri}}, \bibinfo {author}
  {\bibfnamefont {W.}~\bibnamefont {Blanc}}, \bibinfo {author} {\bibfnamefont
  {O.}~\bibnamefont {Legrand}}, \ and\ \bibinfo {author} {\bibfnamefont
  {F.}~\bibnamefont {Mortessagne}},\ }\href@noop {} {\bibfield  {journal}
  {\bibinfo  {journal} {PRE}\ }\textbf {\bibinfo {volume} {85}},\ \bibinfo
  {pages} {047201} (\bibinfo {year} {2012})}\BibitemShut {NoStop}%
\bibitem [{\citenamefont {B\"{a}cker}\ \emph {et~al.}(2008)\citenamefont
  {B\"{a}cker}, \citenamefont {Ketzmerick}, \citenamefont {L\"{o}ck},
  \citenamefont {Robnik}, \citenamefont {Vidmar}, \citenamefont {H\"{o}hmann},
  \citenamefont {Kuhl},\ and\ \citenamefont {St\"{o}ckmann}}]{Backer2008}%
  \BibitemOpen
  \bibfield  {author} {\bibinfo {author} {\bibfnamefont {A.}~\bibnamefont
  {B\"{a}cker}}, \bibinfo {author} {\bibfnamefont {R.}~\bibnamefont
  {Ketzmerick}}, \bibinfo {author} {\bibfnamefont {S.}~\bibnamefont
  {L\"{o}ck}}, \bibinfo {author} {\bibfnamefont {M.}~\bibnamefont {Robnik}},
  \bibinfo {author} {\bibfnamefont {G.}~\bibnamefont {Vidmar}}, \bibinfo
  {author} {\bibfnamefont {R.}~\bibnamefont {H\"{o}hmann}}, \bibinfo {author}
  {\bibfnamefont {U.}~\bibnamefont {Kuhl}}, \ and\ \bibinfo {author}
  {\bibfnamefont {H.~J.}\ \bibnamefont {St\"{o}ckmann}},\ }\href@noop {}
  {\bibfield  {journal} {\bibinfo  {journal} {PRL}\ }\textbf {\bibinfo {volume}
  {100}},\ \bibinfo {pages} {174103} (\bibinfo {year} {2008})}\BibitemShut
  {NoStop}%
\bibitem [{\citenamefont {Lagendijk}\ \emph {et~al.}(2009)\citenamefont
  {Lagendijk}, \citenamefont {van Tiggelen},\ and\ \citenamefont
  {Wiersma}}]{Lagendijk2009}%
  \BibitemOpen
  \bibfield  {author} {\bibinfo {author} {\bibfnamefont {A.}~\bibnamefont
  {Lagendijk}}, \bibinfo {author} {\bibfnamefont {B.}~\bibnamefont {van
  Tiggelen}}, \ and\ \bibinfo {author} {\bibfnamefont {D.~S.}\ \bibnamefont
  {Wiersma}},\ }\href@noop {} {\bibfield  {journal} {\bibinfo  {journal} {Phys.
  Today}\ }\textbf {\bibinfo {volume} {62}},\ \bibinfo {pages} {24} (\bibinfo
  {year} {2009})}\BibitemShut {NoStop}%
\bibitem [{\citenamefont {St\"{o}ckmann}(1999)}]{Stockmann1999}%
  \BibitemOpen
  \bibfield  {author} {\bibinfo {author} {\bibfnamefont {H.-J.}\ \bibnamefont
  {St\"{o}ckmann}},\ }\href@noop {} {\emph {\bibinfo {title} {Quantum chaos: an
  introduction}}},\ \bibinfo {edition} {cambridge}\ ed.\ (\bibinfo  {publisher}
  {Cambridge University Press},\ \bibinfo {year} {1999})\BibitemShut {NoStop}%
\bibitem [{\citenamefont {Ree}\ and\ \citenamefont {Reichl}(1999)}]{Ree1999}%
  \BibitemOpen
  \bibfield  {author} {\bibinfo {author} {\bibfnamefont {S.}~\bibnamefont
  {Ree}}\ and\ \bibinfo {author} {\bibfnamefont {L.~E.}\ \bibnamefont
  {Reichl}},\ }\href@noop {} {\bibfield  {journal} {\bibinfo  {journal} {PRE}\
  }\textbf {\bibinfo {volume} {60}},\ \bibinfo {pages} {1607} (\bibinfo {year}
  {1999})}\BibitemShut {NoStop}%
\bibitem [{\citenamefont {Tureci}\ \emph {et~al.}(2002)\citenamefont {Tureci},
  \citenamefont {Schwefel}, \citenamefont {Stone},\ and\ \citenamefont
  {Narimanov}}]{Tureci2002}%
  \BibitemOpen
  \bibfield  {author} {\bibinfo {author} {\bibfnamefont {H.}~\bibnamefont
  {Tureci}}, \bibinfo {author} {\bibfnamefont {H.}~\bibnamefont {Schwefel}},
  \bibinfo {author} {\bibfnamefont {a.}~\bibnamefont {Stone}}, \ and\ \bibinfo
  {author} {\bibfnamefont {E.}~\bibnamefont {Narimanov}},\ }\href@noop {}
  {\bibfield  {journal} {\bibinfo  {journal} {Opt. Express}\ }\textbf {\bibinfo
  {volume} {10}},\ \bibinfo {pages} {752} (\bibinfo {year} {2002})}\BibitemShut
  {NoStop}%
\bibitem [{\citenamefont {Michel}\ \emph {et~al.}(2007)\citenamefont {Michel},
  \citenamefont {Doya}, \citenamefont {Legrand},\ and\ \citenamefont
  {Mortessagne}}]{Michel2007}%
  \BibitemOpen
  \bibfield  {author} {\bibinfo {author} {\bibfnamefont {C.}~\bibnamefont
  {Michel}}, \bibinfo {author} {\bibfnamefont {V.}~\bibnamefont {Doya}},
  \bibinfo {author} {\bibfnamefont {O.}~\bibnamefont {Legrand}}, \ and\
  \bibinfo {author} {\bibfnamefont {F.}~\bibnamefont {Mortessagne}},\
  }\href@noop {} {\bibfield  {journal} {\bibinfo  {journal} {PRL}\ }\textbf
  {\bibinfo {volume} {99}},\ \bibinfo {pages} {224101} (\bibinfo {year}
  {2007})}\BibitemShut {NoStop}%
\bibitem [{\citenamefont {R\v{o}ska}\ \emph {et~al.}(tted)\citenamefont
  {R\v{o}ska}, \citenamefont {Peleska},\ and\ \citenamefont {Doya}}]{Roska}%
  \BibitemOpen
  \bibfield  {author} {\bibinfo {author} {\bibfnamefont {P.}~\bibnamefont
  {R\v{o}ska}}, \bibinfo {author} {\bibfnamefont {P.}~\bibnamefont {Peleska}},
  \ and\ \bibinfo {author} {\bibfnamefont {V.}~\bibnamefont {Doya}},\
  }\href@noop {} {\  (\bibinfo {year} {submitted})}\BibitemShut {NoStop}%
\bibitem [{\citenamefont {Ghatak}\ and\ \citenamefont
  {Thyagarajan}(1998)}]{Ghatak1998}%
  \BibitemOpen
  \bibfield  {author} {\bibinfo {author} {\bibfnamefont {A.}~\bibnamefont
  {Ghatak}}\ and\ \bibinfo {author} {\bibfnamefont {K.}~\bibnamefont
  {Thyagarajan}},\ }\href@noop {} {\emph {\bibinfo {title} {An Introduction to
  Fiber Optics}}}\ (\bibinfo  {publisher} {Cambride University Press},\
  \bibinfo {address} {Cambridge, UK},\ \bibinfo {year} {1998})\BibitemShut
  {NoStop}%
\bibitem [{\citenamefont {Percival}(1973)}]{Percival1973}%
  \BibitemOpen
  \bibfield  {author} {\bibinfo {author} {\bibfnamefont {I.~C.}\ \bibnamefont
  {Percival}},\ }\href@noop {} {\bibfield  {journal} {\bibinfo  {journal} {J.
  Phys. B At. Mol. Phys.}\ }\textbf {\bibinfo {volume} {6}},\ \bibinfo {pages}
  {L229} (\bibinfo {year} {1973})}\BibitemShut {NoStop}%
\bibitem [{\citenamefont {Husimi}(1940)}]{Husimi1940}%
  \BibitemOpen
  \bibfield  {author} {\bibinfo {author} {\bibfnamefont {K.}~\bibnamefont
  {Husimi}},\ }\href@noop {} {\bibfield  {journal} {\bibinfo  {journal} {Proc.
  Phys. Math. Soc. Jpn.}\ }\textbf {\bibinfo {volume} {22}} (\bibinfo {year}
  {1940})}\BibitemShut {NoStop}%
\bibitem [{\citenamefont {Birkhoff}(1927)}]{Birkhoff1927}%
  \BibitemOpen
  \bibfield  {author} {\bibinfo {author} {\bibfnamefont {G.~D.}\ \bibnamefont
  {Birkhoff}},\ }in\ \href@noop {} {\emph {\bibinfo {booktitle} {Dyn. Syst.}}}\
  (\bibinfo {address} {Cambridge, USA},\ \bibinfo {year} {1927})\BibitemShut
  {NoStop}%
\bibitem [{\citenamefont {Crespi}\ \emph {et~al.}(1993)\citenamefont {Crespi},
  \citenamefont {Perez},\ and\ \citenamefont {Chang}}]{Crespi1993}%
  \BibitemOpen
  \bibfield  {author} {\bibinfo {author} {\bibfnamefont {B.}~\bibnamefont
  {Crespi}}, \bibinfo {author} {\bibfnamefont {G.}~\bibnamefont {Perez}}, \
  and\ \bibinfo {author} {\bibfnamefont {S.-j.}\ \bibnamefont {Chang}},\
  }\href@noop {} {\bibfield  {journal} {\bibinfo  {journal} {PRE}\ }\textbf
  {\bibinfo {volume} {47}} (\bibinfo {year} {1993})}\BibitemShut {NoStop}%
\bibitem [{\citenamefont {B\"{a}cker}\ \emph {et~al.}(2004)\citenamefont
  {B\"{a}cker}, \citenamefont {F\"{u}rstberger},\ and\ \citenamefont
  {Schubert}}]{Backer2004}%
  \BibitemOpen
  \bibfield  {author} {\bibinfo {author} {\bibfnamefont {A.}~\bibnamefont
  {B\"{a}cker}}, \bibinfo {author} {\bibfnamefont {S.}~\bibnamefont
  {F\"{u}rstberger}}, \ and\ \bibinfo {author} {\bibfnamefont {R.}~\bibnamefont
  {Schubert}},\ }\href@noop {} {\bibfield  {journal} {\bibinfo  {journal}
  {PRE}\ }\textbf {\bibinfo {volume} {70}},\ \bibinfo {pages} {036204}
  (\bibinfo {year} {2004})}\BibitemShut {NoStop}%
\bibitem [{\citenamefont {Berry}\ and\ \citenamefont
  {Robnik}(1984)}]{Berry1984}%
  \BibitemOpen
  \bibfield  {author} {\bibinfo {author} {\bibfnamefont {M.~V.}\ \bibnamefont
  {Berry}}\ and\ \bibinfo {author} {\bibfnamefont {M.}~\bibnamefont {Robnik}},\
  }\href@noop {} {\bibfield  {journal} {\bibinfo  {journal} {J. Phys. A. Math.
  Gen.}\ }\textbf {\bibinfo {volume} {17}},\ \bibinfo {pages} {2413} (\bibinfo
  {year} {1984})}\BibitemShut {NoStop}%
\bibitem [{\citenamefont {Rudolf}\ \emph {et~al.}(2012)\citenamefont {Rudolf},
  \citenamefont {Mertig}, \citenamefont {L\"{o}ck},\ and\ \citenamefont
  {B\"{a}cker}}]{Rudolf2012}%
  \BibitemOpen
  \bibfield  {author} {\bibinfo {author} {\bibfnamefont {T.}~\bibnamefont
  {Rudolf}}, \bibinfo {author} {\bibfnamefont {N.}~\bibnamefont {Mertig}},
  \bibinfo {author} {\bibfnamefont {S.}~\bibnamefont {L\"{o}ck}}, \ and\
  \bibinfo {author} {\bibfnamefont {A.}~\bibnamefont {B\"{a}cker}},\
  }\href@noop {} {\bibfield  {journal} {\bibinfo  {journal} {PRE}\ }\textbf
  {\bibinfo {volume} {85}},\ \bibinfo {pages} {036213} (\bibinfo {year}
  {2012})}\BibitemShut {NoStop}%
\bibitem [{\citenamefont {Berry}(1983)}]{Berry1983}%
  \BibitemOpen
  \bibfield  {author} {\bibinfo {author} {\bibfnamefont {M.~V.}\ \bibnamefont
  {Berry}},\ }in\ \href@noop {} {\emph {\bibinfo {booktitle} {Les Houches Lect.
  Ser.}}},\ Vol.~\bibinfo {volume} {36}\ (\bibinfo  {publisher} {North-Holland
  Publishing Compagny},\ \bibinfo {address} {Les Houches},\ \bibinfo {year}
  {1983})\ pp.\ \bibinfo {pages} {171--271}\BibitemShut {NoStop}%
\bibitem [{\citenamefont {Bogomolny}(1988)}]{Bogomolny1988}%
  \BibitemOpen
  \bibfield  {author} {\bibinfo {author} {\bibfnamefont {E.~B.}\ \bibnamefont
  {Bogomolny}},\ }\href@noop {} {\bibfield  {journal} {\bibinfo  {journal}
  {Phys. D Nonlinear Phenom.}\ }\textbf {\bibinfo {volume} {31}},\ \bibinfo
  {pages} {169} (\bibinfo {year} {1988})}\BibitemShut {NoStop}%
\bibitem [{\citenamefont {Kogelnik}\ and\ \citenamefont
  {Li}(1966)}]{Kogelnik1966}%
  \BibitemOpen
  \bibfield  {author} {\bibinfo {author} {\bibfnamefont {H.}~\bibnamefont
  {Kogelnik}}\ and\ \bibinfo {author} {\bibfnamefont {T.}~\bibnamefont {Li}},\
  }\href@noop {} {\bibfield  {journal} {\bibinfo  {journal} {Appl. Opt.}\
  }\textbf {\bibinfo {volume} {5}},\ \bibinfo {pages} {1550} (\bibinfo {year}
  {1966})}\BibitemShut {NoStop}%
\bibitem [{\citenamefont {Feng}\ and\ \citenamefont {Winful}(2001)}]{Feng2001}%
  \BibitemOpen
  \bibfield  {author} {\bibinfo {author} {\bibfnamefont {S.}~\bibnamefont
  {Feng}}\ and\ \bibinfo {author} {\bibfnamefont {H.~G.}\ \bibnamefont
  {Winful}},\ }\href@noop {} {\bibfield  {journal} {\bibinfo  {journal} {Opt.
  Lett.}\ }\textbf {\bibinfo {volume} {26}},\ \bibinfo {pages} {485} (\bibinfo
  {year} {2001})}\BibitemShut {NoStop}%
\bibitem [{\citenamefont {Heller}(1984)}]{Heller1984}%
  \BibitemOpen
  \bibfield  {author} {\bibinfo {author} {\bibfnamefont {E.~J.}\ \bibnamefont
  {Heller}},\ }\href@noop {} {\bibfield  {journal} {\bibinfo  {journal} {PRL}\
  }\textbf {\bibinfo {volume} {53}},\ \bibinfo {pages} {1515} (\bibinfo {year}
  {1984})}\BibitemShut {NoStop}%
\bibitem [{\citenamefont {Pradhan}\ and\ \citenamefont
  {Sridhar}(2000)}]{Pradhan2000}%
  \BibitemOpen
  \bibfield  {author} {\bibinfo {author} {\bibfnamefont {P.}~\bibnamefont
  {Pradhan}}\ and\ \bibinfo {author} {\bibfnamefont {S.}~\bibnamefont
  {Sridhar}},\ }\href@noop {} {\bibfield  {journal} {\bibinfo  {journal} {PRL}\
  }\textbf {\bibinfo {volume} {85}},\ \bibinfo {pages} {2360} (\bibinfo {year}
  {2000})}\BibitemShut {NoStop}%
\bibitem [{\citenamefont {Kudrolli}\ \emph {et~al.}(1995)\citenamefont
  {Kudrolli}, \citenamefont {Kidambi},\ and\ \citenamefont
  {Sridhar}}]{Kudrolli1995}%
  \BibitemOpen
  \bibfield  {author} {\bibinfo {author} {\bibfnamefont {A.}~\bibnamefont
  {Kudrolli}}, \bibinfo {author} {\bibfnamefont {V.}~\bibnamefont {Kidambi}}, \
  and\ \bibinfo {author} {\bibfnamefont {S.}~\bibnamefont {Sridhar}},\
  }\href@noop {} {\bibfield  {journal} {\bibinfo  {journal} {PRL}\ }\textbf
  {\bibinfo {volume} {75}},\ \bibinfo {pages} {818} (\bibinfo {year}
  {1995})}\BibitemShut {NoStop}%
\bibitem [{\citenamefont {Pradhan}\ and\ \citenamefont
  {Sridhar}(2002)}]{Pradhan2002a}%
  \BibitemOpen
  \bibfield  {author} {\bibinfo {author} {\bibfnamefont {P.}~\bibnamefont
  {Pradhan}}\ and\ \bibinfo {author} {\bibfnamefont {S.}~\bibnamefont
  {Sridhar}},\ }\href@noop {} {\bibfield  {journal} {\bibinfo  {journal}
  {Pramana - J. Phys.}\ }\textbf {\bibinfo {volume} {58}},\ \bibinfo {pages}
  {333} (\bibinfo {year} {2002})}\BibitemShut {NoStop}%
\bibitem [{\citenamefont {Prigodin}\ and\ \citenamefont
  {Altshuler}(1998)}]{Prigodin1998}%
  \BibitemOpen
  \bibfield  {author} {\bibinfo {author} {\bibfnamefont {V.}~\bibnamefont
  {Prigodin}}\ and\ \bibinfo {author} {\bibfnamefont {B.~L.}\ \bibnamefont
  {Altshuler}},\ }\href@noop {} {\bibfield  {journal} {\bibinfo  {journal}
  {PRL}\ }\textbf {\bibinfo {volume} {80}},\ \bibinfo {pages} {4} (\bibinfo
  {year} {1998})}\BibitemShut {NoStop}%
\bibitem [{\citenamefont {Feit}\ and\ \citenamefont {Fleck}(1978)}]{Feit1978}%
  \BibitemOpen
  \bibfield  {author} {\bibinfo {author} {\bibfnamefont {M.~D.}\ \bibnamefont
  {Feit}}\ and\ \bibinfo {author} {\bibfnamefont {J.~a.}\ \bibnamefont
  {Fleck}},\ }\href@noop {} {\bibfield  {journal} {\bibinfo  {journal} {Appl.
  Opt.}\ }\textbf {\bibinfo {volume} {17}},\ \bibinfo {pages} {3990} (\bibinfo
  {year} {1978})}\BibitemShut {NoStop}%
\bibitem [{\citenamefont {Feit}\ and\ \citenamefont {Fleck}(1980)}]{Feit1980}%
  \BibitemOpen
  \bibfield  {author} {\bibinfo {author} {\bibfnamefont {M.~D.}\ \bibnamefont
  {Feit}}\ and\ \bibinfo {author} {\bibfnamefont {J.~a.}\ \bibnamefont
  {Fleck}},\ }\href@noop {} {\bibfield  {journal} {\bibinfo  {journal} {Appl.
  Opt.}\ }\textbf {\bibinfo {volume} {19}},\ \bibinfo {pages} {2240} (\bibinfo
  {year} {1980})}\BibitemShut {NoStop}%
\bibitem [{\citenamefont {Gutzwiller}(1990)}]{Gutzwiller1990}%
  \BibitemOpen
  \bibfield  {author} {\bibinfo {author} {\bibfnamefont {M.~C.}\ \bibnamefont
  {Gutzwiller}},\ }\href@noop {} {\emph {\bibinfo {title} {Chaos in Classical
  and Quantum Mechanics}}}\ (\bibinfo  {publisher} {Springer-Verlag},\ \bibinfo
  {address} {New York},\ \bibinfo {year} {1990})\BibitemShut {NoStop}%
\bibitem [{\citenamefont {Balian}\ and\ \citenamefont
  {Bloch}(1971)}]{Balian1971}%
  \BibitemOpen
  \bibfield  {author} {\bibinfo {author} {\bibfnamefont {R.}~\bibnamefont
  {Balian}}\ and\ \bibinfo {author} {\bibfnamefont {C.}~\bibnamefont {Bloch}},\
  }\href@noop {} {\bibfield  {journal} {\bibinfo  {journal} {Ann. Phys.}\
  }\textbf {\bibinfo {volume} {63}} (\bibinfo {year} {1971})}\BibitemShut
  {NoStop}%
\bibitem [{\citenamefont {Lebental}\ \emph {et~al.}(2007)\citenamefont
  {Lebental}, \citenamefont {Djellali}, \citenamefont {Arnaud}, \citenamefont
  {Lauret}, \citenamefont {Zyss}, \citenamefont {Dubertrand}, \citenamefont
  {Schmit},\ and\ \citenamefont {Bogomolny}}]{Lebental2007a}%
  \BibitemOpen
  \bibfield  {author} {\bibinfo {author} {\bibfnamefont {M.}~\bibnamefont
  {Lebental}}, \bibinfo {author} {\bibfnamefont {N.}~\bibnamefont {Djellali}},
  \bibinfo {author} {\bibfnamefont {C.}~\bibnamefont {Arnaud}}, \bibinfo
  {author} {\bibfnamefont {J.~S.}\ \bibnamefont {Lauret}}, \bibinfo {author}
  {\bibfnamefont {J.}~\bibnamefont {Zyss}}, \bibinfo {author} {\bibfnamefont
  {R.}~\bibnamefont {Dubertrand}}, \bibinfo {author} {\bibfnamefont
  {C.}~\bibnamefont {Schmit}}, \ and\ \bibinfo {author} {\bibfnamefont
  {E.}~\bibnamefont {Bogomolny}},\ }\href@noop {} {\bibfield  {journal}
  {\bibinfo  {journal} {PRA}\ }\textbf {\bibinfo {volume} {76}},\ \bibinfo
  {pages} {023830} (\bibinfo {year} {2007})}\BibitemShut {NoStop}%
\bibitem [{\citenamefont {Aschieri}\ \emph {et~al.}(2011)\citenamefont
  {Aschieri}, \citenamefont {Garnier}, \citenamefont {Michel}, \citenamefont
  {Doya},\ and\ \citenamefont {Picozzi}}]{Aschieri2011}%
  \BibitemOpen
  \bibfield  {author} {\bibinfo {author} {\bibfnamefont {P.}~\bibnamefont
  {Aschieri}}, \bibinfo {author} {\bibfnamefont {J.}~\bibnamefont {Garnier}},
  \bibinfo {author} {\bibfnamefont {C.}~\bibnamefont {Michel}}, \bibinfo
  {author} {\bibfnamefont {V.}~\bibnamefont {Doya}}, \ and\ \bibinfo {author}
  {\bibfnamefont {A.}~\bibnamefont {Picozzi}},\ }\href@noop {} {\bibfield
  {journal} {\bibinfo  {journal} {PRA}\ }\textbf {\bibinfo {volume} {83}},\
  \bibinfo {pages} {033838} (\bibinfo {year} {2011})}\BibitemShut {NoStop}%
\end{thebibliography}%

\end{document}